\documentclass[prb,twocolumn,groupaddress,superscriptaddress,showpacs]{revtex4}

\usepackage{graphicx}
\usepackage{amssymb}
\usepackage[usenames]{color}
\usepackage{enumerate}

\usepackage{amsmath}
\usepackage{natbib}
\usepackage{nicefrac}
\usepackage{multirow}
\usepackage{epstopdf}
\usepackage{float}
\begin{document}

\title{Multiple Mode Torsional Oscillator Studies \\
and Evidence for Supersolidity in Bulk $^4$He}

\author{A. Eyal}
\affiliation{Department of Physics, LASSP, Cornell, Ithaca
NY 14850}

\author{X. Mi}
\affiliation{current address: Department of Physics, Princeton, NJ 08544}

\author{A. V. Talanov}

\affiliation{Department of Physics, LASSP, Cornell, Ithaca
NY 14850}

\author{J. D. Reppy}

\affiliation{Department of Physics, LASSP, Cornell, Ithaca
NY 14850}

\date{\today}

\begin{abstract}

The discovery by Kim and Chan (KC) \cite{KC,KCScience} of an anomalous decrease in the period of torsional oscillators (TO) containing samples of solid $^4$He at temperatures below 0.2 K was initially interpreted as a superfluid-like decoupling of a fraction of the solid moment of inertia from the TO. These experiments appeared to confirm the thirty-year-old theoretical prediction by Chester \cite{Chester}, Andreev \cite{Andreev}, and Leggett \cite{Leggett} of the existence of a low-temperature Bose-condensed supersolid phase in solid $^4$He. The initial results of KC lead to a flurry of experimental and theoretical activity and their results for bulk $^4$He samples were soon confirmed in a number of laboratories \cite{ReppyConfirm2006,Kojima,Kubota,Shirahama}. The early excitement in this field, however, has been tempered by the realization that an anomaly in the shear modulus of solid helium $^4$He may explain most if not all of the period shifts observed in the early TO experiments. In principle, it is possible to distinguish experimentally between shear modulus induced period shifts and period shifts resulting from other physical mechanisms through the use of multiple frequency torsional oscillators. In this paper we shall discuss our recent results on bulk solid $^4$He with double and triple mode TOs, using both open cylindrical cells and annular cells. In these experiments, we have observed a small frequency independent contribution to the period shift signal such as might be expected in the presence of a supersolid phase. Since the interplay between the elastic properties of the solid and the mechanics of TOs can be subtle and depends on the specific design of the TO, we shall discuss in detail the mechanics of the TOs employed in our measurements applying both analytic and finite element methods (FEM) for the analysis. 

\end{abstract}

\pacs {67.80 bd.}
\maketitle

	\section{Introduction} 

The idea of a possible existence of a supersolid state, a new phase of matter, excited a large number of experimental and theoretical work over the past forty five years, especially in the past decade. Such a state is assumed to present properties of superflow in a, most likely bosonic, solid. It was first suggested to exist based on theoretical grounds by Andreev \cite{Andreev}, Chester \cite{Chester} and Leggett \cite{Leggett} in the late 1960s and early 1970s. As liquid $^4$He turns into superfluid, and because it is a bosonic material with a low mass and weak interatomic interactions, solid $^4$He was a natural candidate in the search for supersolidity.
Torsional oscillators (TO) have been the chief instrument in the investigation of this possible supersolid state of $^4$He. The use of TOs in the search for this state follows a suggestion by Leggett in 1970 \cite{Leggett} that the appearance of the supersolid state would result in an anomalous superfluid-like decoupling of a portion of the moment of inertia of the solid $^4$He sample from the TO in a manner analogous to the decoupling of the superfluid fraction of liquid $^4$He as observed by Andronikashvili \cite{Andronikashvili} in his early TO experiments. Such a reduction of the moment of inertia upon entering the superfluid or supersolid state was termed by Leggett to be a non-classical-rotational-inertia (NCRI). The NCRI reduction in the moment of inertia is then expected to begin to increase with decreasing temperature below an onset temperature and lead to an observable decrease in the period of a TO containing a supersolid sample.  
The first attempt to observe the proposed supersolid state employing a TO, at Bell Labs \cite{Bishop}, gave negative results and it was concluded that a supersolid signal, if present, was below the detection level of few parts in 10$^5$.  
However, subsequent TO experiments by E. Kim and H.W.M. Chan (KC) gave positive results and the anomalous decrease in the TO period or NCRI expected for the supersolid state was observed. The first observations were obtained for solid $^4$He samples contained in porous Vycor glass \cite{KC}. Later, the NCRI effect was observed by KC \cite{KCScience} for bulk samples of solid $^4$He. KC demonstrated in a series of experiments that the magnitude of the NCRI was velocity dependent, being reduced at high wall velocities, and also sensitive to the impurity concentration of $^3$He. These results for the bulk solid were soon replicated in a number of different laboratories \cite{ReppyConfirm2006,Kojima,Kubota,Shirahama} and were initially interpreted as evidence for a supersolid phase in solid $^4$He.
Over time, the supersolid interpretation of the early TO data has been brought into question by the increasing awareness of the importance of the significant role that an anomaly in the elastic constants of solid $^4$He can play in contributing to the period shifts observed in the TO experiments. Following the earlier work at Bell Labs by Paalanen, Bishop, and Dail \cite{Paalanen}, Day and Beamish (DB) \cite{Beamish} at the University of Alberta made a detailed study of the temperature dependence of the shear modulus of solid $^4$He. DB found an anomalous decrease in the shear modulus of the solid as the temperature was raised from low temperatures below 40 mK to above 250 mK. This is the same temperature range over which the NCRI phenomenon has been observed in TO experiments. In their early work, DB report fractional changes in the shear modulus of up to 10 percent. Depending on the design of a TO, such variations in the helium shear modulus with temperature can lead to period shift signals with a temperature dependence and magnitude similar to those seen by KC and others. One example for this is the careful repetition \cite{ChanNoVycor} of the early solid $^4$He in Vycor \cite{KC} experiment by the Penn state group, in which an upper limit of 2x10$^{-5}$ was set for supersolid in Vycor. The period shifts in the original experiment probably arose from an increase of shear modulus $\mu$ of a thin layer of bulk solid $^4$He in the TO cells. These results were also confirmed by double TO measurements \cite{ReppyVycor} of solid helium in Vycor, where all of the apparent signal was attributed to elastic effects. 
Therefore, the current consensus in the field is that a supersolid phase has probably not been observed experimentally in solid $^4$He so far. However, both recent theoretical predictions \cite{Anderson} and simulations \cite{Pollet,PolletGB} don't rule out the existence of some type of superflow in the solid. 
In an attempt to overcome the problems raised by the interplay between the mechanics of TOs and the shear modulus anomaly, we have employed multiple frequency TOs in a strategy to discriminate between period shifts arising from the elastic anomaly and those due to the existence of a possible supersolid. This approach has been successful in that in our most recent experiments with three TOs of different design, we have observed the existence of a sample dependent frequency independent contribution to the observed period shifts. The magnitude of the observed frequency-independent contributions, when normalized by the period shift measured upon freezing of the sample, ranges from 0.4$\times$10$^{-4}$ to 1.5$\times$10$^{-4}$. Such frequency-independent signals are possible candidates of the period shifts expected for the supersolid phase.
In this paper we give a detailed discussion of these results including a careful consideration of the influence of the elastic anomaly on the mechanics of our individual TOs. We shall also include a discussion of previous multiple frequency TO experiments \cite{Kojima,KojimaPrb} of Kojima's group at Rutgers University and the more recent experiment \cite{Cowan} by Cowan's group at Royal-Holloway. In the discussion of the interplay between temperature dependent changes in the solid $^4$He elastic modulus and the TOs we have employed both analytic calculations and a finite element methods (FEM) approach for our analysis. An important conclusion from our analysis is that any frequency independent contribution arising from the elastic anomaly can be strongly suppressed by careful design and the remaining elastic contribution will be frequency dependent, chiefly varying as the square of the TO frequency.

		\subsection{Extended details of the search} 

			\subsubsection{Brief early history} 

The early measurements of Kim and Chan \cite{KC,KCScience} showed both the period shift and the velocity dependence that were expected to be present. Their signals range between 0.75\% and 2.5\% of the total moment of inertia of the solid sample. These numbers, $\Delta P/\Delta P_F$, where $\Delta P$ is the period change attributed to the supersolid transition, and $\Delta P_F$ the change in period upon freezing of the solid sample, can be defined as fractional period shifts (FPS). That is analogous to the superfluid density ratio, or what is called NCRIF in many of the papers on the subject. 
The signals observed varied from sample to sample, with wide variation in signal size from the several percent level to the detection level of the measurements. This extreme dependence on samples, in addition to dependence on cell geometry, gave early indications, that were not immediately fully appreciated, that the experimental situation was not simple.

			\subsubsection{Unexpected results}
As other groups started repeating and extending the KC results, more confusing results were observed. One of the first of these were Rittner and Reppy's annealing experiments \cite{ReppyConfirm2006}, which gave early indications that the subject was more complicated than had first appeared. Annealing the samples for which signals were observed would result in significant reduction of the signals. On the other hand, quench cooling of the samples, which generated stress and disorder, resulted in increased signal size. 
Double mode oscillator experiments \cite{Kojima} revealed frequency dependent signals, which were also inconsistent with a simple supersolid scenario. Flow measurements, performed at first by Hallock’s group \cite{Hallock} and later by Chan \cite{ChanFlow} and Beamish \cite{BeamishFlow} revealed mass flow which disappeared below about 80mK, where the TO signals were largest. Rotating oscillator experiments, performed by the groups of Kubota \cite{KubotaRotating}, Shirahama \cite{ShirahamaRotating} and Kim \cite{Choi} showed a rotation dependence of the signals, with a velocity dependence different than that obtained by the TO velocity measurements. On the other hand, recent DC rotation measurements by the Golov group \cite{GolovRotating} show no such dependence. 
On the theoretical level, Path Integral Monte Carlo simulations performed by Pollet et al. \cite{Pollet} show no superflow in defect free crystals, but do show superfluidity of grain boundaries, and a Luttinger liquid in the cores of some dislocations. Anderson \cite{AndersonSupersolids}, on the other hand, demonstrated superlow when investigating the Bose Hubbard model, and suggested that $^4$He would exhibit the same behaviour. 

			\subsubsection{Elastic properties of solid $^4$He}
\begin{figure}[]
\includegraphics[width=3.5in]{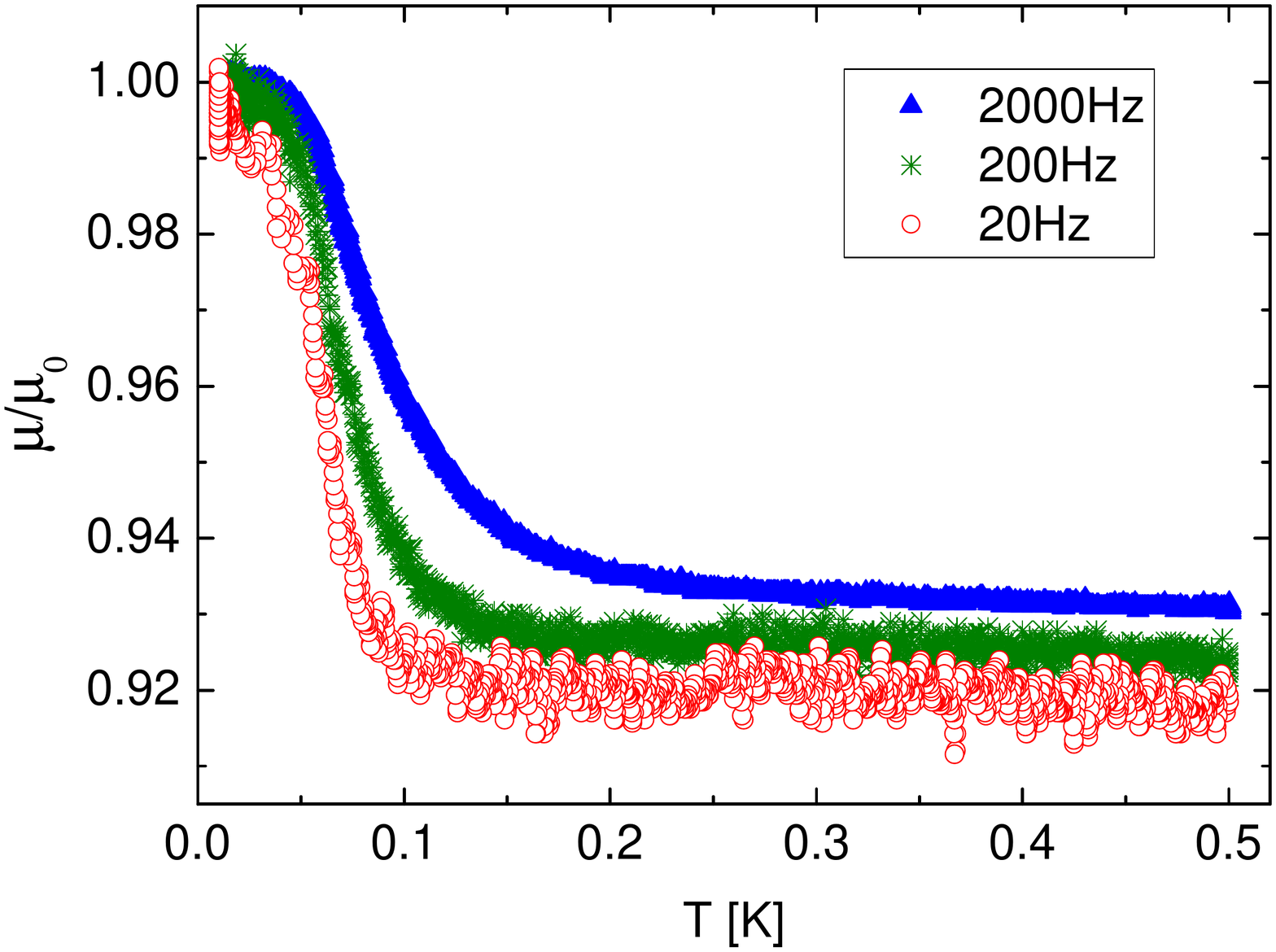} {}
\caption{(Color Online) Normalized shear modulus versus temperature of a polycrystalline sample of $^4$He, for three different excitation frequencies. $\mu_0$=1.5x10$^8$ dyn/cm$^2$. Data supplied by John Beamish.}
\label{fig:BeamishShearModulus}
\end{figure}

Day and Beamish discovered in 2007 \cite{Beamish} an anomalous softening with increasing temperature of the shear modulus of solid helium over the same temperature range as the supposed supersolid, thus introducing complications to the subject. The anomalous period shifts observed in torsional oscillator experiments can be partially explained by the elastic anomaly, and therefore are not necessarily a signature of a supersolid state. Given the importance of elastic effects in the temperature dependent period of TOs employed in the supersolid search, we will start with a brief summary of the elastic properties of solid $^4$He.
In Fig. \ref{fig:BeamishShearModulus} the shear modulus data, supplied by John Beamish, for a polycrystalline sample, are plotted as a function of temperature. The $^4$He sample in these measurements was contained between two closely spaced parallel piezoelectric plates. One plate was excited in low frequency transverse motion. The second piezoelectric plate served to detect the strain induced in the solid by the motion of the active plate. There are three distinguishable temperature regions seen in these data. There is a high temperature region, above roughly 200 to 300 mK, where the shear modulus shows only a weak variation with temperature. In the region between 200 mK and 50 mK, however, there is a remarkable increase in the magnitude of the shear modulus to a maximum value achieved below 50 mK, in another temperature independent region. Although the decrease in the shear modulus for a polycrystalline sample as shown in Fig. \ref{fig:BeamishShearModulus} is only 8\% of the low temperature value, much larger reductions have been observed in other experiments with single crystal samples. For example, in the giant plasticity experiments performed by Balibar's group at the ENS in Paris \cite{BalibarGiantPlasticity}, reductions in the shear modulus in excess of 50\% were observed for single crystals between the low and high temperature regions.

\begin{figure}[]
\includegraphics[width=3in]{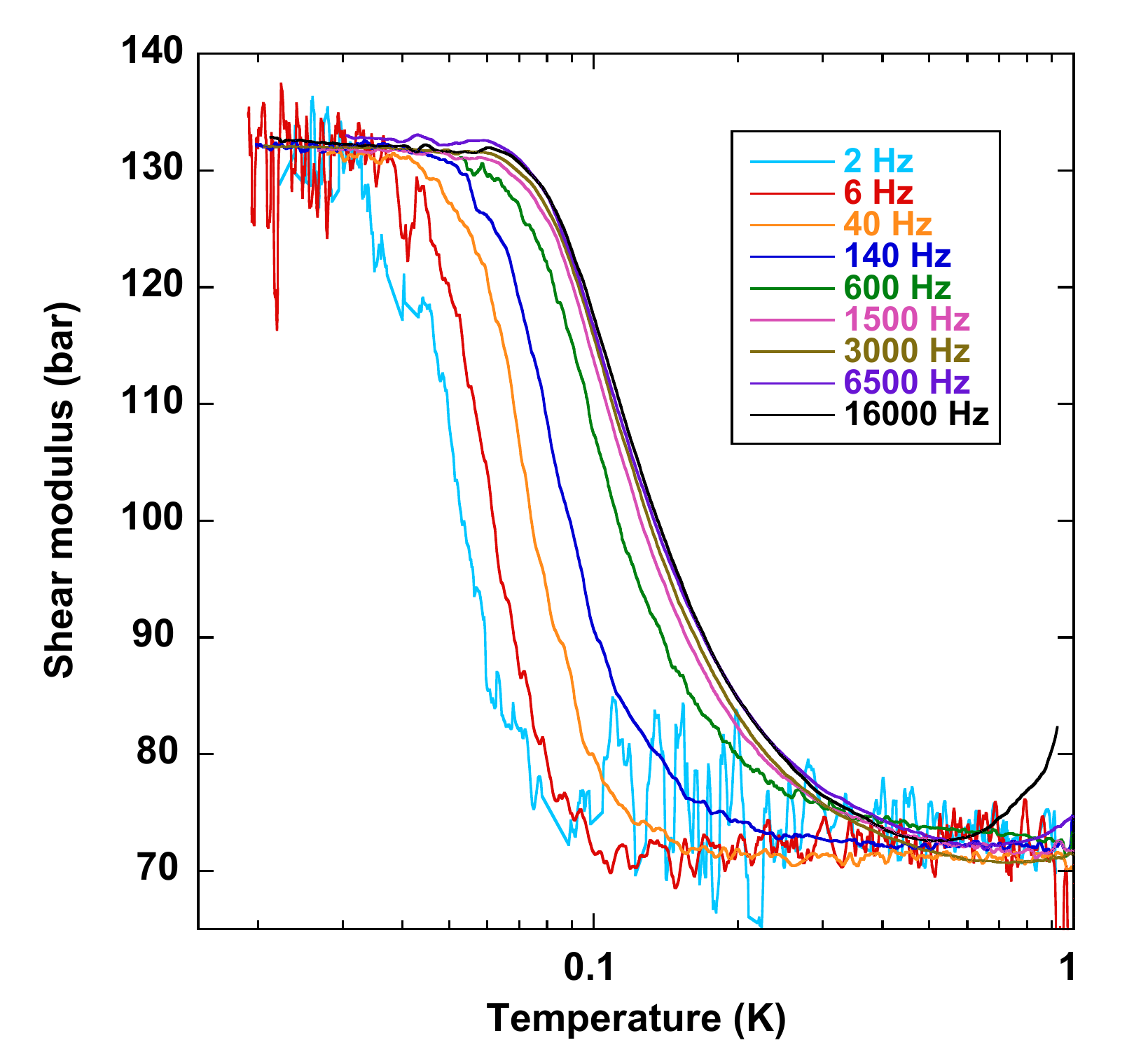} {}
\caption{(Color Online) Shear modulus versus temperature for several different excitation frequencies for a single crystal of $^4$He. Data taken from \cite{BalibarDislocationsPrb}.}
\label{fig:BalibarShearModulus}
\end{figure}

DB demonstrated a sensitivity to $^3$He concentration similar to that observed by KC. They also investigated the frequency dependence of the anomaly as well as the dependence on the strain level determined by the amplitude of the active plate. Fig. \ref{fig:BalibarShearModulus} shows recent $^4$He shear modulus data, from the Paris group of Balibar \cite{BalibarDislocationsPrb}, as a function of temperature for frequencies ranging from 2 to 2x10$^4$ Hz. At the lowest temperature, and also in the region between 400 and 500 mK, there is little dependence on frequency as compared to the region where the shear modulus is declining rapidly with increasing temperature, where a frequency dependence is clearly evident.

The current understanding of the phenomena of this elastic anomaly is based on the temperature dependent pinning of dislocation lines by the $^3$He impurities. Given a sufficient concentration of $^3$He atoms in the $^4$He solid, the dislocation lines will be strongly pinned at low temperatures, however, as the temperature is raised, the dislocations become unpinned and mobile as the pinning $^3$He atoms “evaporate” from the lines. The increased mobility of the dislocation lines with unpinning leads to a reduction in the shear modulus of the solid $^4$He. Once the $^3$He has completely evaporated, the shear modulus becomes again, as in the low temperature region, relatively independent of temperature. At considerably higher temperatures the mobility of the dislocations is restricted by damping from thermal excitations in the solid. Increasing the stress on the samples also results in “tearing” the $^3$He atoms away from the dislocation lines, hence creating an apparent velocity dependence of the signals. For a comprehensive discussion of the rather complex problem of the motion of dislocation lines in solid $^4$He, see the review article by Balibar et al. \cite{BalibarReview}.
In our work, the frequency range of the multiple frequency TOs is restricted to a range between 600 to 2500 Hz, much smaller than that shown in Fig. \ref{fig:BalibarShearModulus}. In the analysis of the data, we will be emphasizing the changes in TO period for changes in temperatures between the low temperature region below 50 mK and the high temperature region between 300 and 500 mK, thus avoiding the region where the shear modulus is strongly frequency dependent.

			\subsection{Importance of the elastic effects}

Over the past several years there has been a gradual emergence of an appreciation of the significance of the interplay between elastic properties of the solid and the mechanics of TOs. These explain much, if not all, of what had been seen in the TO experiments. 
As shown in the DB plot (Fig. \ref{fig:BeamishShearModulus}), $\mu$ decreases with increasing temperature, this leads to an increase in the period of a TO when the shear modulus plays a role in the stiffness of the TO, either in the fill line or in the body of the sample cell, thus mimicking the reduction of the period with increasing temperature expected for a supersolid. These TO period shifts can arise from either one of the following:

\begin{enumerate}
  \item Acceleration of the solid helium in the cell \cite{Reppy_JLTP_Review2012}.
  \item The “Maris effect” - the counter-stress of solid on the cell walls \cite{Maris_Effect}.
  \item Twisting of the cell walls. 
  \item The fill line effect - stiffening of the solid $^4$He in the torsion rods \cite{Beamish_TorsionRodEffect}.
  \item The dissipative dynamics of the solid \cite{Graf_Glassy_DoubleTO}.
  \item Coupling different mobile parts of the cell via the solid helium.
\end{enumerate}

There are two main approaches to calculating these different contributions: 

\begin{enumerate}[(a)] 
  \item The analytic approach, in which simple geometries are used to give approximations and upper limits of the observed contributions. These show a frequency squared dependence of the elastic contributions inside the cell, suggesting the use of multiple frequency TOs as a tool to distinguish between these effects and a possible supersolid signal. 
  \item Finite Element Method (FEM) modeling can give more accurate results for specific cell geometries. In this approach as well, where only elastic effects are being considered, plots of FEM FPS vs. f$^2$ show no frequency independent contributions for cells in which the torsion rod effect is not present or significant.
\end{enumerate}

In the appendix, we provide a more detailed discussion and calculations of the known elastic effects of solid $^4$He for one of our TOs, using both an analytical approach and a FEM. Based on these calculations, the estimated FPS values for each elastic effect at the two resonance frequencies are shown in Fig. \ref{fig:FEMFPS}. It is clear that, for our cell, the only elastic effect on the order of $10^{-4}$ arises from the acceleration of solid $^4$He, with a FPS proportional to f$^2$, the square of the TO frequency. 

A main concern is that in single frequency oscillators it is not possible to distinguish between a possible real supersolid and an elastic contribution. In recent work by the Penn state group \cite{ChanRigid} a very rigid cell was employed in attempt to decrease as much as possible all elastic effects. Two different rigid oscillator designs were used to examine samples at pressures between 34 and 60 bar with rim velocities between 11 and 19 $\mu$m/sec. An NCRI signal of 4x10$^{-6}$ was obtained for one cell while the other showed a value of 2x10$^{-5}$.

			\subsection{Solution: Multiple mode TOs}

The effect of the helium's shear modulus change on the TO's period leads to the problem as to how to distinguish a genuine supersolid signal from one induced by changes in the elastic modulus of the $^4$He sample. One needs to consider the dynamics of the TO and the interplay between the TO mechanics and the elastic properties of the sample. One conclusion to be emphasized is that if one hopes to identify a true supersolid signal then any potential candidate should be put to the multiple frequency test in order to eliminate elastic contributions. It is critical in the design of a TO that the significant elastic effects should be understood and reduced to a minimum in order to facilitate the analysis of the observed FPS. In such a case there will be only two contributions to the observed signal - a frequency squared elastic term and a frequency independent supersolid signal. A double frequency TO would then be enough to distinguish between the two. 

For the simple case of a double oscillator, placed on a vibration isolator (VI) satisfying the condition $I_\text{VI} \gg I_\text{cell}$ and $I_\text{VI} \gg I_\text{dummy}$, for example, the influence of the VI on the resonance periods of the TO is small and one can find the eigen-modes of the system by solving
\begin{multline}
 \left( \begin{array}{cc}
-\omega ^2 I_c +k_c & -k_c \\
-k_c & -\omega ^2 I_d + k_c + k_d\end{array} \right) 
 \left( \begin{array}{c}
\theta _c \\
\theta _d\end{array} \right) =
\left( \begin{array}{c}
0 \\
0\end{array} \right)
\end{multline}

Here $k_c$ and $k_d$ are the torsion constants of the rods of the cell and dummy piece, $I$ are the moment of inertia, and $\theta$ are the angular displacements.

Therefore the two resonance angular frequencies can be well approximated by
\begin{multline}
  \omega_\pm^2 = \biggl( \frac{I_\text{c} \left( k_\text{c} +  k_\text{d} \right) +I_\text{d} k_\text{c}}{2 I_\text{c} I_\text{d}}\biggr) \times \\
\biggl( 1 \pm \sqrt{1 - \frac{4 I_\text{c} I_\text{d} k_\text{c} k_\text{d}}{(I_\text{c} \left( k_\text{c} +  k_\text{d} \right) +I_\text{d} k_\text{c})^2}} \biggr)
\end{multline}

The sensitivity to changes in torque for a TO can be determined experimentally from the period shifts observed upon freezing of the sample or estimated analytically from the equations of motion for the TO. 
In the case of a single mode TO, when the moment of inertia of the solid sample, $I_{He}$, is much less than $I_c$, the moment of inertia of the sample cell, the sensitivity is given by the approximate expression:
\begin{equation}
\frac{\Delta p}{p}=-\frac{1}{2} \frac{I_{He}}{I_c}
\end{equation}
In the case of a two mode TO, the expression for the sensitivity has an additional factor, and is given by the expression:
\begin{equation}
\biggl( \frac{\Delta p}{p} \biggr)_{\pm}=-\frac{1}{2} \frac{I_{He}}{I_c} \frac{\omega_c^2-\omega_{\mp}^2}{\omega_{\pm}^2-\omega_{\mp}^2}
\end{equation}
where $\omega_c^2=\frac{k_c}{I_c}$.
For our open cylindrical cell we get a reduction in the sensitivity of about a factor of 2 compared to the case of a single
mode oscillator. Even with that reduction, our sensitivity to changes in the moment of inertia of the helium is around 0.01, comparable to the numbers reported by the Chan group for their high sensitivity long path length experiments \cite{ChanLongPath}.

For the case of triple mode oscillators, or when taking the VI into account, there is a closed form for the frequencies, and they can be easily determined using root finding algorithms. 

	\section{Experiments}
		\subsection{General apparatus and torsional oscillator designs}

\begin{figure}[]
\includegraphics[width=3.5in]{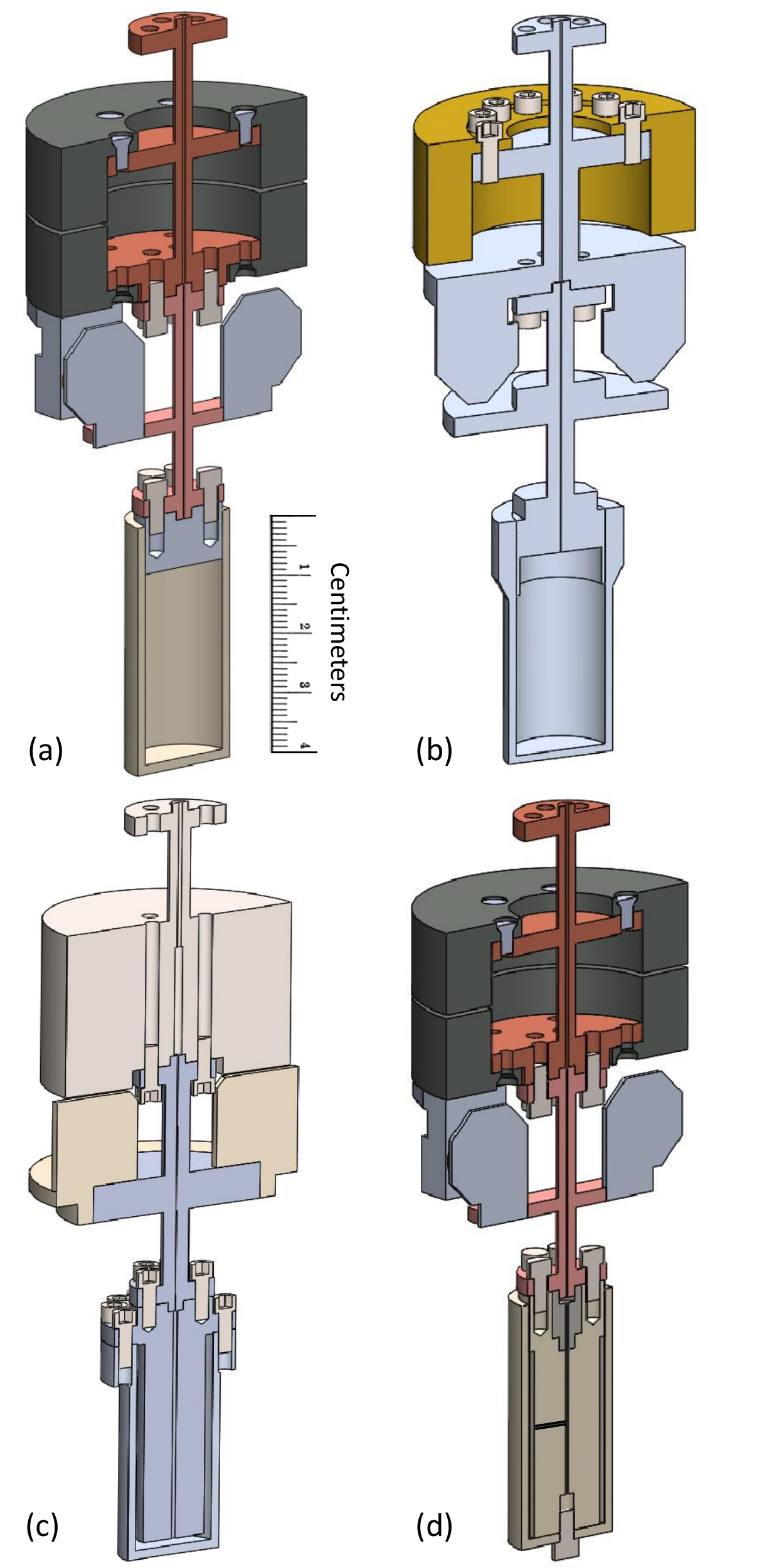} {}
\caption{(Color Online) (a) A double TO design with a cylindrical sample space. (b) A triple TO with an open cylinder sample space. (c) A double TO design with an annular sample space. (d) A double TO design with an annular sample space that has a non-simply connected geometry. All TOs are attached to vibration isolators.}
\label{fig:Cells}
\end{figure}

In the work reported here, we have employed TOs of two different designs; an open cylinder design for the sample geometry and an annular sample geometry. There have also been several iterations of our basic design making minor changes in an attempt to improve the quality of our data. In Fig. \ref{fig:Cells} we show schematics of four of our TOs. (a) and (b) are an open cylinder design and in (c) the solid $^4$He sample is contained in an annular region. The TO in (d) is only given as an example of another cell we used, and the results of it are not reported in this paper.  

The frequencies and mass loading sensitivities for these cells are:

\begin{tabular}{l || c | c | c | c }

   & (a) & (b) & (c) & (d) \\
\hline
$f_+$ [Hz] & 2073.6 & 2129.3 & 1543.0 & 1970.2 \\
$f_{mid}$ [Hz] &   & 1336.9 &  &  \\
$f_-$ [Hz] & 758.6  & 706.1 & 646.0 & 717.6\\
$\Delta p_{F+}$ [$\mu$ sec] & 5.18  & 2.68 & 1.69  & 1.48 \\
$\Delta p_{F\,mid}$ [$\mu$ sec] &   & 1.80 &   &  \\
$\Delta p_{F-}$ [$\mu$ sec] & 13.48  & 8.96 & 2.75 & 5.07 \\
\end{tabular}
\\

\begin{figure}[]
\includegraphics[width=3.5in]{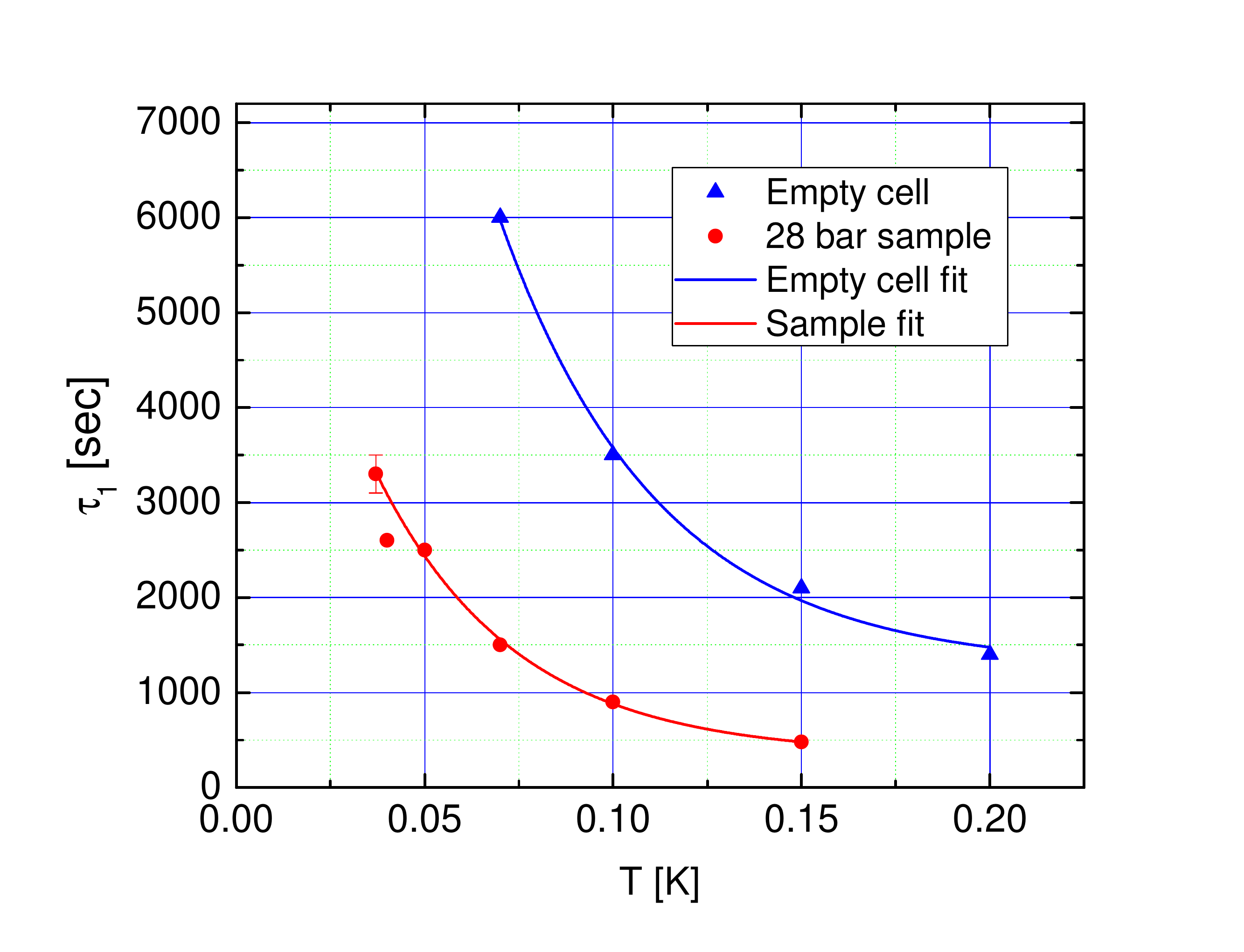} {}
\caption{(Color Online) Relaxation times vs. temperature for both an empty cell and a cell full of solid $^4$He. Data taken from the high mode of the triple mode TO.}
\label{fig:taus}
\end{figure}

The cells were driven and detected using parallel plate capacitors. The detection was mostly done with two capacitors connected in parallel, placed on opposite sides on the cell, to reduce floppy mode vibrations. In some of our later designs one capacitor plate was placed on the dummy piece, while the other was placed on the vibration isolator, to further reduce vibrational effects on the detection. The moving electrode was biased with 200VDC. 
For most of our measurements the modes were excited simultaneously with identical maximum rim velocities, usually around 4$\mu$m/sec for each mode. We also made several measurements with different velocities and with constant driving amplitudes, rather than at constant detection amplitude, as well as several measurements where only a single mode was excited. 
The general course of measurements was the following: prior to the formation of a solid $^4$He sample we made careful background measurements of the period $p_\pm$ of the oscillator with vacuum in the cell. This was done either by slowly sweeping the temperature, or by heating / cooling in steps, waiting for the period to relax to a constant value at each temperature step. 
The relaxation times observed at each temperature step are strongly temperature dependent as shown if Fig. \ref{fig:taus} where the relaxation times observed for our triple TO are shown.  
The large increase in the relaxation time with decreasing temperature is suggestive of thermal activation. 

Measurements of the dependence of the period of the empty cell on the velocity / acceleration were also made. After the empty cell data was obtained, we formed polycrystalline samples by the blocked-capillary method from commercial $^4$He gas, having a nominal 0.3 ppm $^3$He impurity level. We then repeated the temperature and velocity altering measurements we had performed for the empty cell. 
The period change $\Delta p(T)$ is the deviation of the period of the TO from the empty cell background, which is matched at 300mK to the full cell's values. The 300mK temperature was chosen since at this temperature the shear modulus is relatively independent of both frequency and temperature. 

		\subsection{Expectations from a supersolid state}

The supersolid state is assumed to be a Bose-condensed state in a solid sample of $^4$He and as such it should present properties resembling those of superfluids. One of these expected basic properties is a clear phase transition temperature. Specifically, if a solid sample contained in a torsional oscillator is cooled below the supersolid transition temperature, the period of the oscillator is expected to decrease, depending on size of the supersolid fraction. That period drop should be independent of the frequency of the oscillator, assuming the velocity of the cell is below the critical velocity. Critical velocity is yet another basic property of the supersolid. Increasing the velocity of the cell below the transition temperature should cause a reduction in the period change. These properties should also hold in the case of a lower dimensional superflow in the solid, such as superflow of grain boundaries or dislocation cores \cite{Pollet}. 

		\subsection{Our results}
We performed experiments using both open cylindrical cells and annular cells, with both double and triple mode oscillators, made of several materials and observed a qualitatively similar behaviour in all cases. All our samples were grown using the blocked capillary method, with sample pressures ranging between 26bar and 34bar.
The mass loading - the period change upon filling the cell with solid - was measured both upon growing the samples and when melting them at a constant temperature, usually of 0.5K. The pressure dependence of the cell was taken into account when comparing the two. 

			\subsubsection{Double mode open cylinder}
One of the reasons we chose to use an open cylinder was to increase the size of the small expected signal. A cross section of the double open cylinder TO used in our measurements appears in Fig. \ref{fig:Cells} (a). The cell itself was made out of Mg, the torsion rods were BeCu, and as was the dummy piece. The TO was mounted on a double vibration isolator with Cu torsion rods and Pb moments of inertia. The outer Mg wall of the cell was tightly connected to an Al TO base with a 36 threads per inch screw joint. The threads were coated with TRA bond epoxy to provide a leak tight seal. The torsion rod, in turn, was attached to the Al base. The inner radius of the cell was 0.794cm, the outer radius 0.953cm, the inner height 3.251cm and the width of the bottom 0.159cm. The Al plate glued to the cell was 0.794cm thick. The torsion rods had a diameter around 0.37cm. The rod close to the cell was 1.05cm long, and the one between the dummy moment of inertia and the vibration isolator (VI) was 1.71cm long. Steel inserts were epoxied or soldered into holes in the copper VI to provide a hard surface for the lead o-ring seals between the TO and the VI.
In the design of our TO’s we have taken care to minimize the inner diameter of the cell fill line to reduce the influence of changes in the shear modulus of the solid in the fill line on the frequency of the TO. In the current design, the cell was filled through a 0.1mm ID x 0.25 OD CuNi tube, which passed through a 0.71 mm diameter hole drilled along the central axis of the torsion rods. The void between the outer diameter of the fill line and the hole drilled in the torsion rod is filled with 1266 Stycast epoxy to exclude the possibility of solid $^4$He in this region. 
The different moments of inertia and torsion constants were:\\

\begin{figure}[]
\includegraphics[width=3.5in]{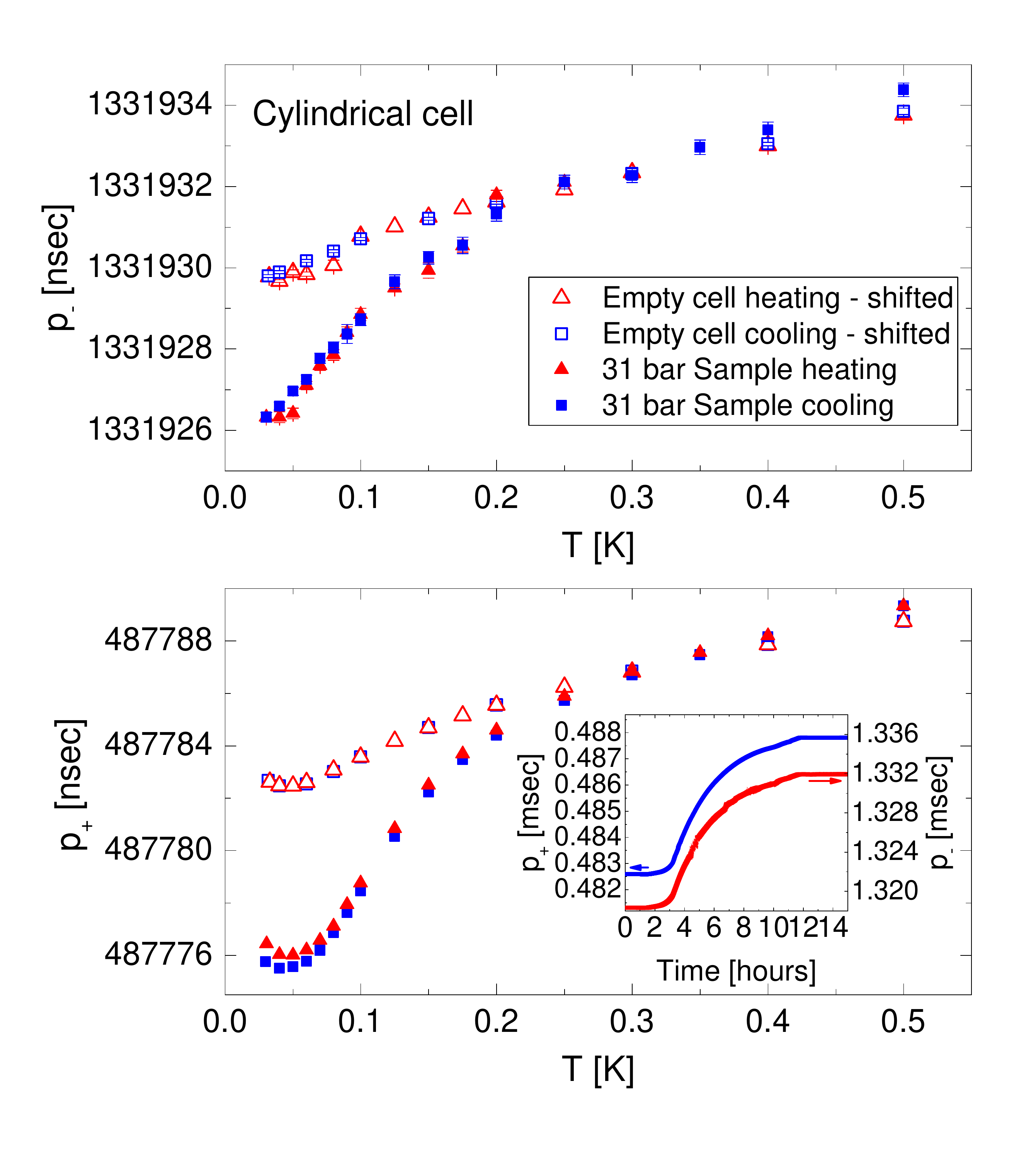} {}
\caption{(Color Online) Period versus temperature for a 31 bar sample. The top panel showing the low 750 Hz mode and bottom one the high 2050Hz mode. Empty cell data points were shifted to match the full cell at 300mK. Both modes were driven with a constant maximal velocity of v=4$\mu$m/sec. The inset shows the period of the modes upon forming the sample.}
\label{fig:DoubleP}
\end{figure}

\begin{tabular}{ l ||  l }

  Torsion constant  [dyn cm]  & Moment of inertia [g cm$^2$]\\
\hline
  $K_{cell}$= 9.29E+08  & $I_{cell}$= 9.0 \\
  $K_{dummy}$= 5.26E+08 & $I_{dummy}$= 13.7 \\
   $K_{VI1}$= 3.59E+08 & $I_{VI1}$= 957.8 \\
 $K_{VI2}$= 8.62E+08  & $I_{VI2}$= 886.4 \\
  &  $I_{He}$= 0.405 \\
\end{tabular}
\\
\\
For this TO the electrode holders for the detection were mounted on the vibration isolator. The capacitances of the detection were 2.3pF and 2.5pF. 
The low mode frequency of the empty cell was 758.60Hz and 2073.61Hz for the high mode. The corresponding frequency shifts upon solidification of the sample were 7.68Hz and 22.04Hz respectively. These give periods shifts of $\Delta p_{F-}$=13.48 $\mu$ sec and $\Delta p_{F+}$=5.18 $\mu$ sec of mass loading for the different modes, which translate to $\Delta p_F /p$=0.01 for both modes. These numbers require an accuracy of only a few parts of a mHz to detect 10$^{-4}$ signals. The frequency stability for our modes was less than 0.06mHz for the low mode, and less than 0.02mHz for the high one. The Qs were around 10$^5$ for the high mode and 2.5x10$^5$ for the low mode at low temperatures. 
We also measured the pressure dependence of the period of the cell. The period changed by 1 nsec for the high mode and 2.2 nsec for the low mode for each 1 bar change. These give corrections around 0.5\% for the mass loadings. 
Fig. \ref{fig:DoubleP} shows the period versus temperature of the two modes of this TO both upon cooling and heating for a sample with p=31bar. The sample pressure was estimated from the melting temperature of 1.8K, taken upon slow heating of the sample. The full markers show the equilibrium values of the period of the TO containing the sample at the different temperatures, and the empty markers are the period of the empty cell, shifted to match the values of the full cell at 300mK. The difference between those two data sets is the period shift $\Delta p$. All points were taken using a constant maximal velocity of 4$\mu$m/sec for each mode. 
The inset shows the period shifts $\Delta p_F$ upon growing a sample. 

\begin{figure}[]
\includegraphics[width=3.5in]{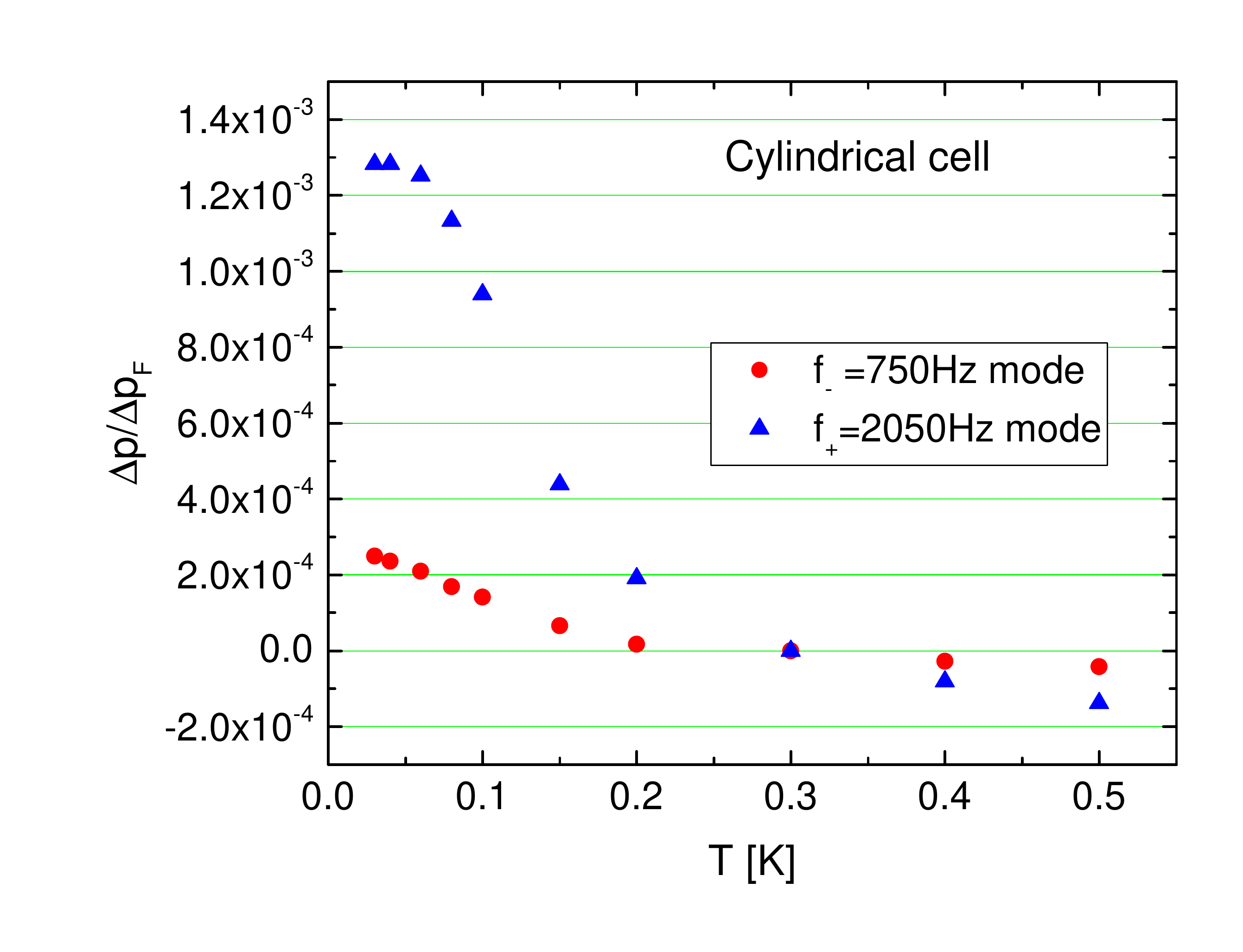} {}
\caption{(Color Online) Fractional period change versus temperature for a 31 bar sample. T=300mK was set as zero.}
\label{fig:DoubleFPSvsT}
\end{figure}

For this particular sample the mass loadings were $\Delta p_{F-}$=13741.19nsec and $\Delta p_{F+}$=5294.43nsec, from reducing the cell pressure at 500mK. Dividing $\Delta p$ taken from the data of fig. \ref{fig:DoubleP} by these mass loadings gives us the temperature dependent fractional period shifts (FPS). These are plotted in Fig. \ref{fig:DoubleFPSvsT} for both modes. If the observed signals were attributed entirely to supersolidity, we would expect the FPS to be frequency-independent and depend only on temperature. If, on the other hand, there is no supersolid, and the signal comes entirely from the elastic effect arising chiefly from the acceleration of the $^4$He solid (as our cell is designed to suppress other elastic effects), then,
according to the assumptions discussed in the appendix, the signal would depend on the frequency of the mode with an f$^2$ dependence. The results actually show both. 

\begin{figure}[]
\includegraphics[width=3.2in]{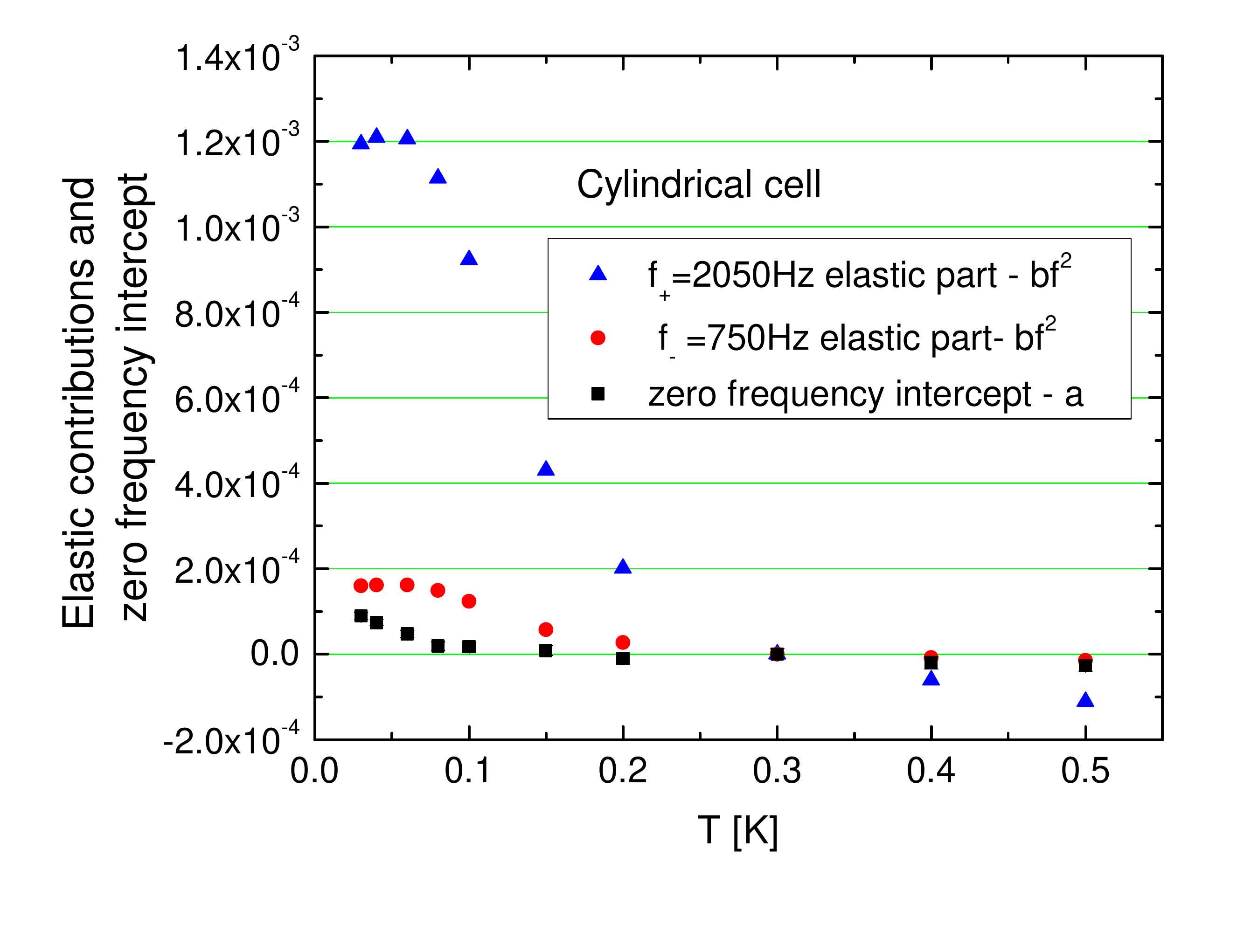} {}
\caption{(Color Online) The different contributions to the signal, $a(T)$ and $b(T) f_\pm ^2$, defined in the relation $\Delta p_\pm (T) / \Delta P_{\text{F}\pm} = a(T) + b(T) f_\pm ^2$, for a 31 bar sample.}
\label{fig:DoubleElastic}
\end{figure}

We can therefore decompose the measured $\Delta p/\Delta p_F$ of the two different modes into these contributions: For each mode, the FPS can be written as $\Delta p_\pm (T) / \Delta p_{F\pm} = a(T) + b(T) f_\pm ^2$, where $a(T)$ is the temperature-dependent supersolid (or other frequency independent) contribution to the FPS. The elastic contribution to the FPS is $b(T) f_\pm^2$. In Fig. \ref{fig:DoubleElastic}, we plot both the elastic contributions to the FPS and the frequency independent one as a function of temperature. It can be seen that the elastic contribution continues to change above $T$ = 0.2 K. This feature is consistent with the continuing change in $\mu$ above 0.2 K, and is seen in previous TO experiments that are dominated by elastic effects \cite{Mi_PRL2012,Mi_QFS2013}. The frequency independent term, however, seems to level out above around 100mK. 

\begin{figure}[]
\includegraphics[width=3.5in]{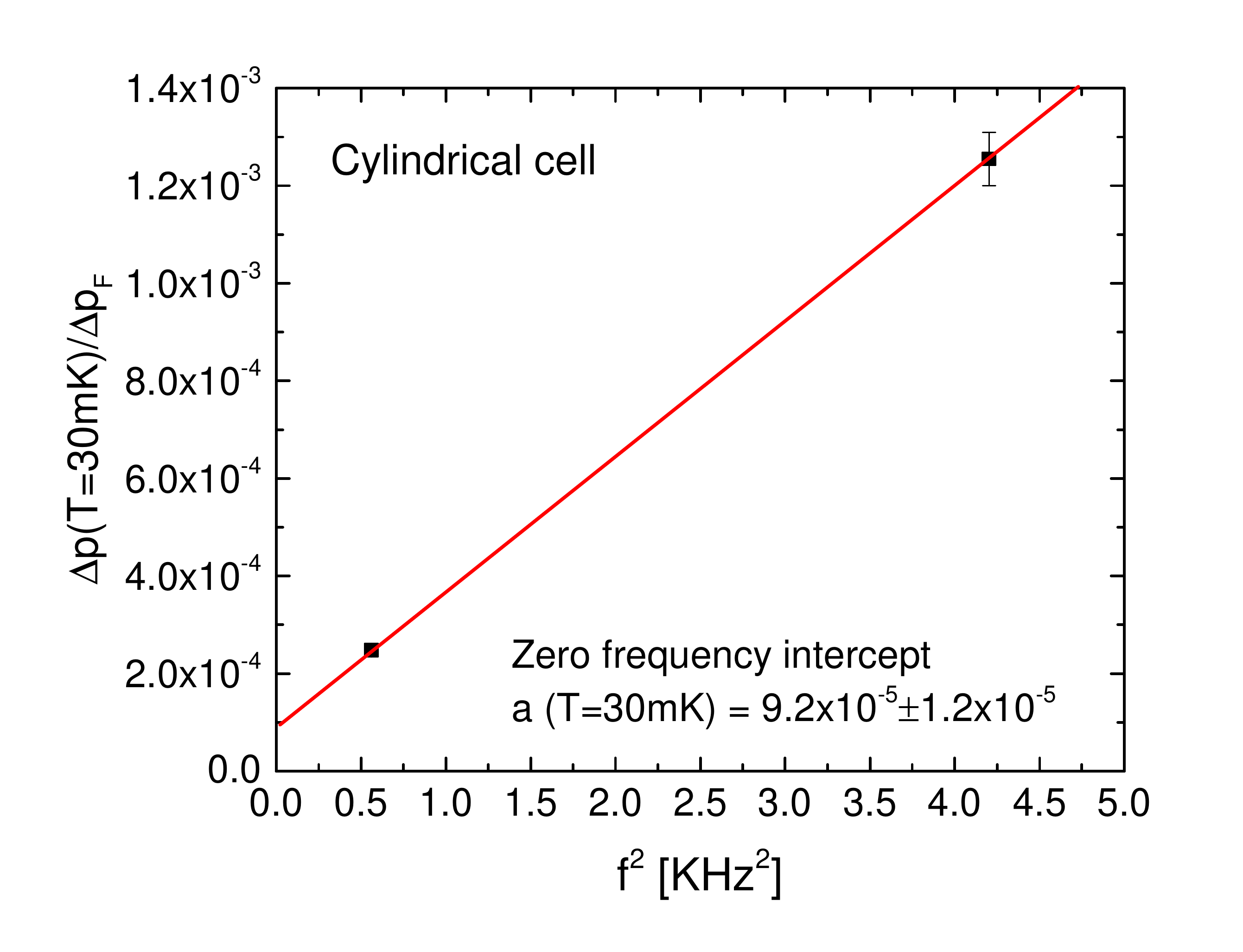} {}
\caption{(Color Online) The fractional period shift for the two modes versus their frequency squared.}
\label{fig:DoubleFPSvsf2}
\end{figure}

Another way to present the data is shown in Fig. \ref{fig:DoubleFPSvsf2}, which shows a plot of the FPSs vs f$^2$. Here $\Delta p$ is taken as the period shift between 30mK and 300mK, where the empty cell data matching was done. These temperatures were also chosen as their shear modulus is only slowly varying with temperature and has a relatively small frequency dependence. 
The zero frequency intercept in this case is the part of the signal that does not depend on frequency, and therefore is not an elastic contribution, but possibly a supersolid term.  

\begin{figure}[]
\includegraphics[width=4in]{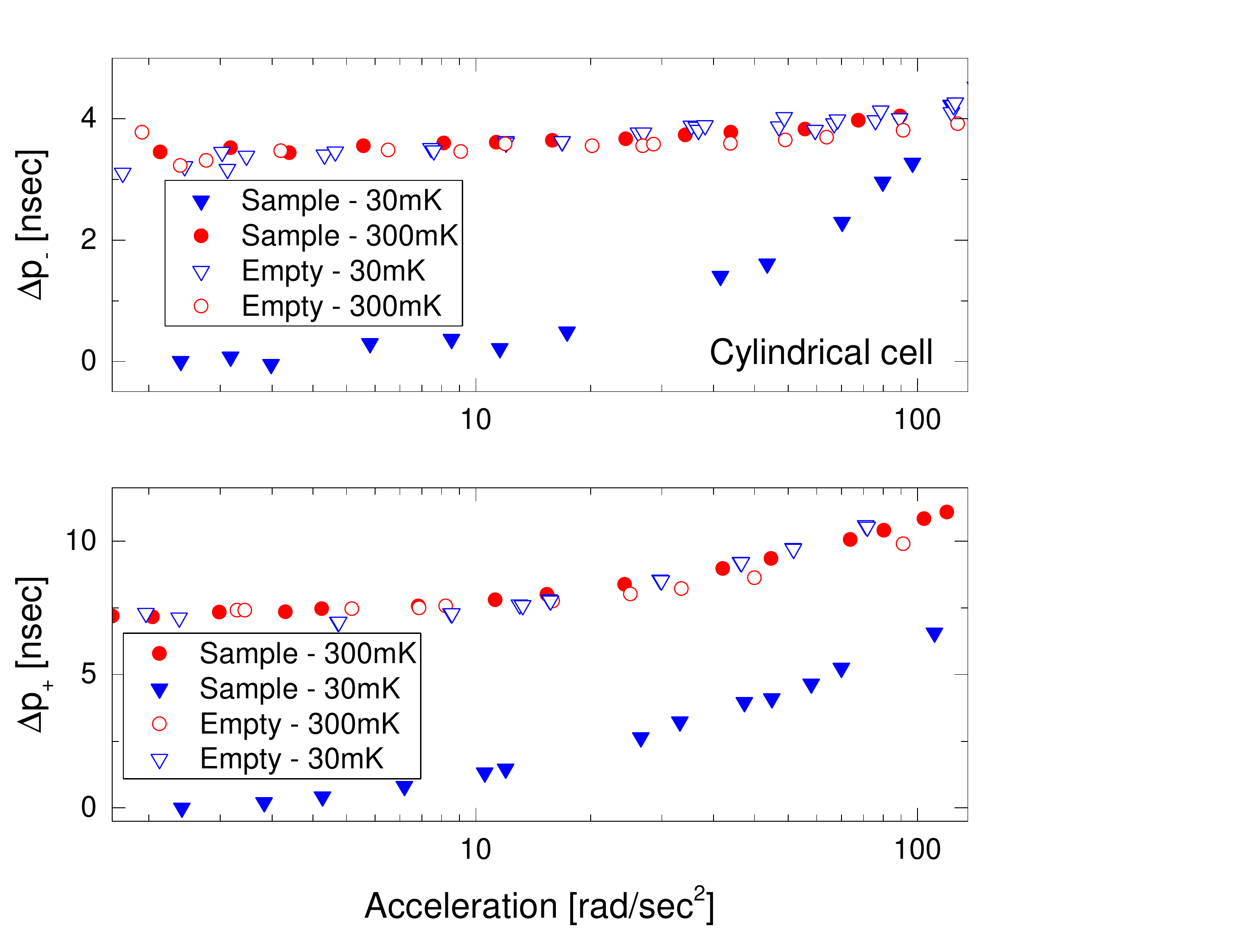} {}
\caption{(Color Online) Acceleration dependence of the period of the different modes for a 31 bar sample. The data points were shifted to give the measured FPS at the accelerations used for the temperature dependent data shown in figures \ref{fig:DoubleP}-\ref{fig:DoubleFPSvsf2}. The empty cell data points were shifted to match those of the full cell at 300mK to enable a clearer comparison of their acceleration dependence.}
\label{fig:DoublePvsacc}
\end{figure}

The modes were usually driven simultaneously for our experiments, but we repeated the measurements with the modes driven separately upon heating and cooling, and got the same results, with a small 0.67 nsec difference for the high mode. The high mode's error bar in Fig. \ref{fig:DoubleFPSvsf2} arises from both that difference and the difference between the heating and cooling curves. 
The acceleration or velocity dependence of the sample was carefully measured. We measured both the empty and full cell's velocity dependence ranging over more than 2 orders of magnitude, with the modes driven separately, at two different temperatures. The first one being the low temperature of 30mK, and the second one was 300mK.
The sample's 300mK data's acceleration dependence is the same as that of the empty cell both at 30mK and at 300mK - which agrees well with choosing the 300mK temperature as the point where the low temperature effects no longer appear. Fig. \ref{fig:DoublePvsacc} shows the acceleration dependence of the two modes at these two temperatures, as well as the empty cell data at those temperatures.

\begin{figure}[]
\includegraphics[width=3.5in]{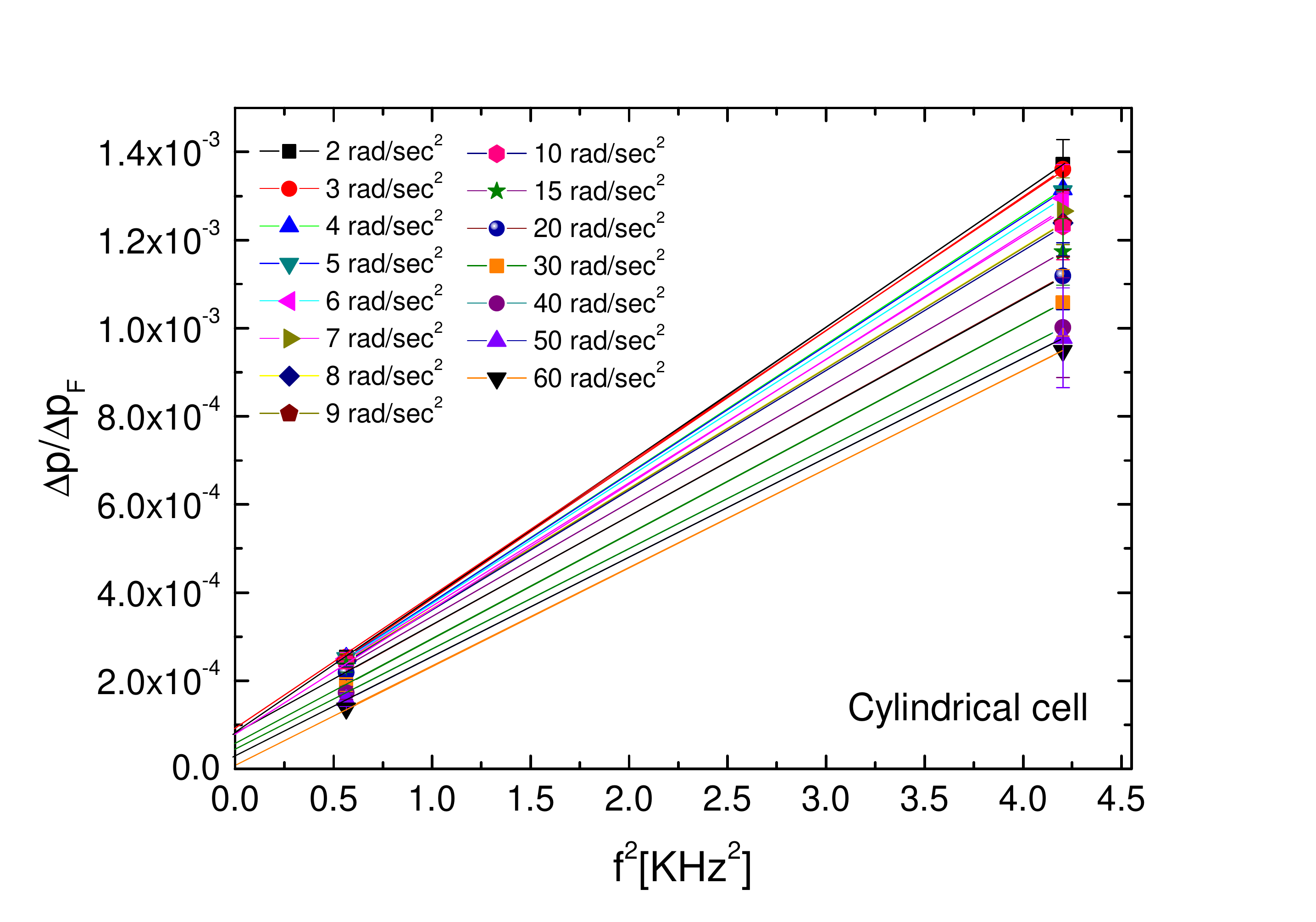} {}
\caption{(Color Online) Acceleration dependence of the FPS for the two modes vs. their frequency squared. The intercept of the straight lines with the f=0Hz line gives the frequency independent term.}
\label{fig:DoubleFPSvsf2acc}
\end{figure}

\begin{figure}[]
\includegraphics[width=3.5in]{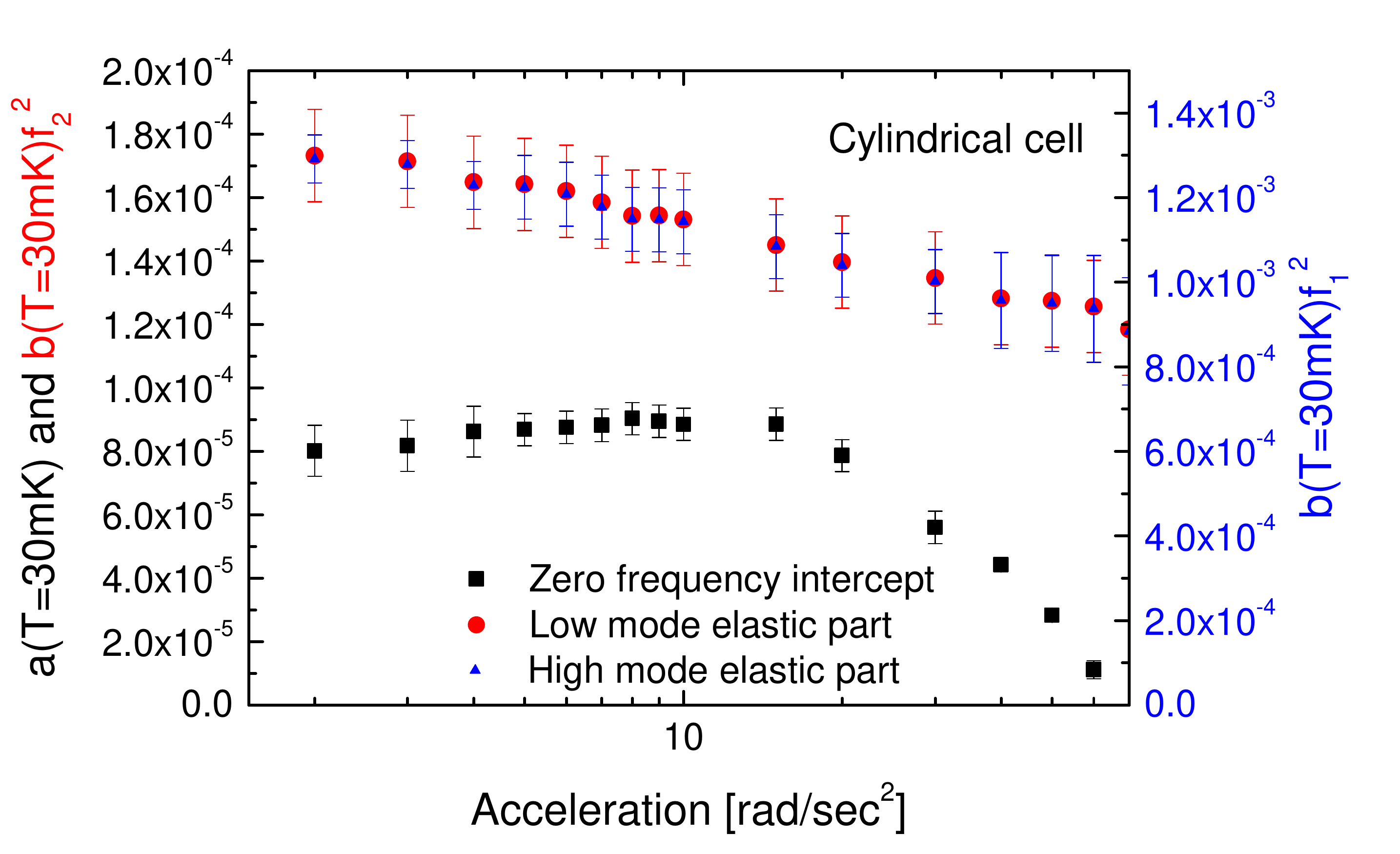} {}
\caption{(Color Online) The zero frequency intercept and the elastic terms, taken from fitting the data plotted in Fig. \ref{fig:DoubleFPSvsf2acc}, versus acceleration. The error bars come from matching the sample signal with the empty cell acceleration dependence.}
\label{fig:DoubleElasticvsAcc}
\end{figure}

Fig. \ref{fig:DoubleFPSvsf2acc} shows the acceleration dependence of the FPSs vs. f$^2$, and Fig. \ref{fig:DoubleElasticvsAcc} shows the acceleration dependence of the zero frequency intercepts taken from Fig  \ref{fig:DoubleFPSvsf2acc}.  

As can be seen in this figure, the zero frequency intercept appears to not depend on the acceleration up to 15rad/sec$^2$ where it starts dropping, and reaches zero around 60rad/sec$^2$. The elastic parts of the signal, on the other hand, show a continuous drop in the signal size with acceleration.  In terms of shear stress, 15rad/sec$^2$ are 0.05 Pa, or a strain of 3x10$^{-9}$, which, comparing to shear modulus measurement experiments \cite{Beamish2010}, is below the 10$^{-7}$ critical strain reported, and is well into the fully pinned region for the $^3$He atoms on the dislocations at 30mK \cite{KimHysteretic}. 
We did not measure above 100rad/sec$^2$ as that made irreversible changes to the frequency. 100rad/sec$^2$ are 0.3Pa, which, according to a recent paper by Kim \cite{KimHysteretic}, is in the hysteretic region for the dislocation behaviour. At high enough accelerations even the frequency of the empty cell was altered, and the temperature of the cell could be affected.

			\subsubsection{Triple mode open cylinder}
In order to test the validity of our assumption, that in our double mode TO’s there are only two significant contribution: the frequency independent one, and an f$^2$ dependent one, we constructed a triple compound oscillator. 
A schematic cross section of the TO can be seen in Fig. \ref{fig:Cells} (b). The entire TO was made out of aluminum 6061, and the torsion rods $K_{cell}$ and $K_{dummy1}$ were heat treated at 413 degrees Celsius for 3 hours, followed by a slow cool down to room temperature. The triple oscillator was placed on a vibration isolator (VI) with a brass piece as a moment of inertia, and an aluminum torsion rod. The Al rod, $K_{dummy2}$, connecting the TO to the VI was 1.55cm long with a 6.35 mm diameter and had a 0.343mm diameter hole drilled through its center for filling the cell, and the two other rods (both 1.27cm long with a 4.57mm diameter) had a 0.71mm diameter hole filled with Stycast 1266 through which a tube with an inner diameter of 0.1mm was inserted, and served to fill the cell. The bottom and top torsion rods were connected by screws and so was the brass piece. The cell itself was an open cylinder that was sealed with a TRA-BOND 2151 on the threads connecting it to the end of the torsion rod. The height of the inner part of the cell was 3.3 cm, the inner radius was 0.79 cm, the wall's and bottom part's thickness was 0.13 cm, and the top of the cell was 0.95 cm thick.
The torsion constants and moments of inertia of the different parts of the cell were:\\

\begin{tabular}{ l ||  l }

  Torsion constant  [dyn cm]  & Moment of inertia [g cm$^2$]\\
\hline
  $K_{cell}$= 9.11E+08  & $I_{cell}$= 11.4 \\
  $K_{dummy1}$= 9.11E+08 & $I_{dummy1}$= 18.3  \\
   $K_{dummy2}$= 2.77E+09 & $I_{dummy2}$= 54.6 \\
 $K_{VI}$= 4.03E+08 & $I_{VI}$= 805.8 \\
  &  $I_{He}$= 0.41 \\
\end{tabular}
\\
\\

The measured empty cell frequencies below 1 K were $f_1$=2129.3Hz, $f_2$=1336.9Hz, $f_3$=706.1Hz, and the Qs were all in the several 10$^5$ range. These translate to the following resonant periods: $p_1$=0.470msec, $p_2$=0.748msec, and $p_3$=1.416msec.  
Our frequency stability for the different modes was less than 0.008nsec for the high mode, 0.034nsec for the middle mode, and 0.14nsec for the low mode. 
The pressure dependence of the cell, as measured at 5K by changing the cell pressure between 0 and 60 bar was $\Delta p/\Delta p_F$=2x$10^{-4}$/bar for all three modes. Here $\Delta p_F$ is the measured mass loading of the cell, which depends on the molar volume of the sample, and hence on the growth temperature. For one set of data, for example for a 28 bar sample, these were measured to be $\Delta p_{1F}$=2682nsec,$\Delta p_{2F}$=1802nsec,$\Delta p_{3F}$=8956nsec. Correcting these values by the pressure dependence changes them only by about 0.5\% of the $\Delta p_{F}$ of the different modes. 
The moving electrode for this oscillator was attached to $I_{dummy2}$. We used two detection capacitors, both of 2.5pF. 

\begin{figure}[]
\includegraphics[width=3.5in]{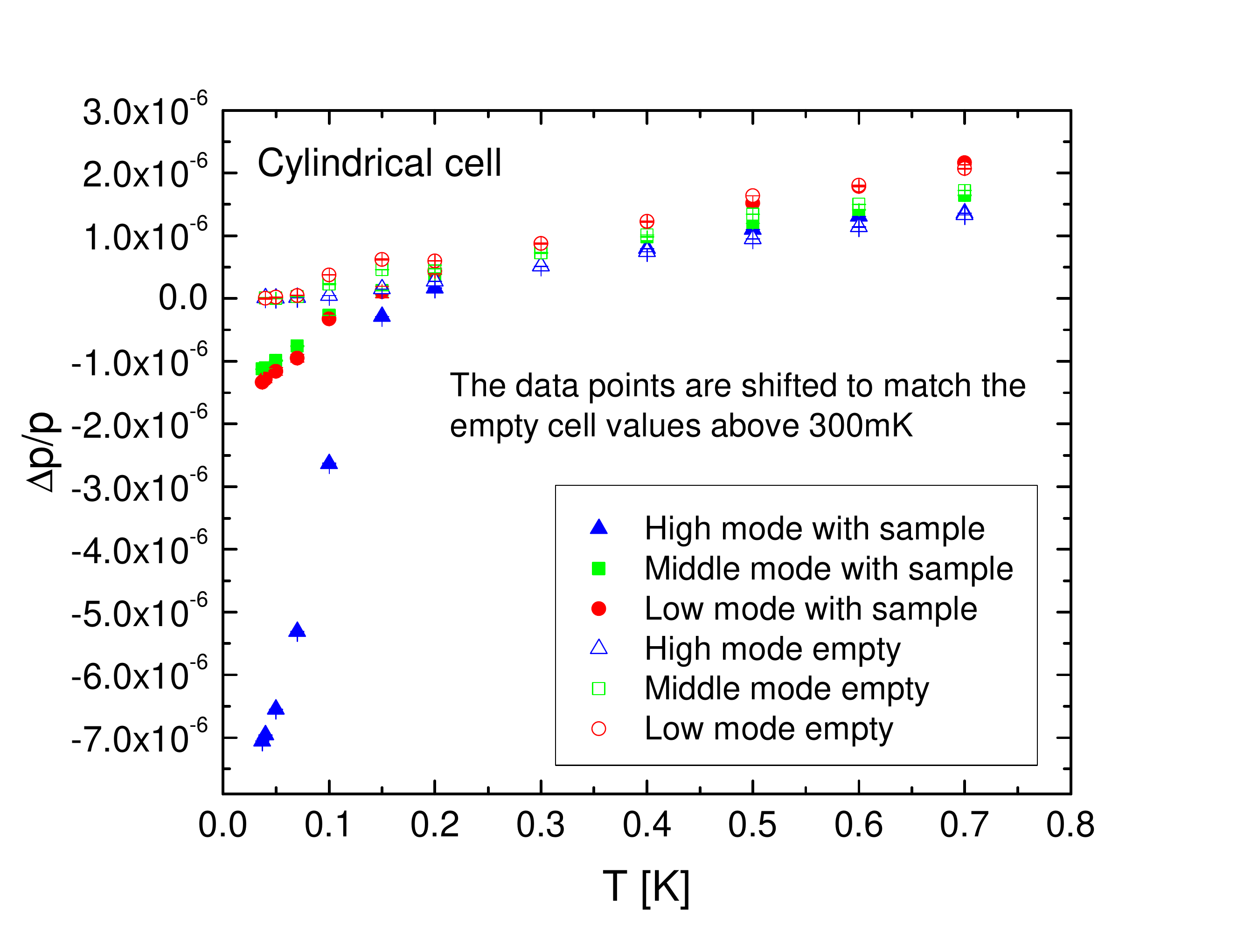} {}
\caption{(Color Online) Relative period change for a sample with pressure around 28bar. All three modes and their empty cell backgrounds are shown.}
\label{fig:TripleP}
\end{figure}

We grew 5 different samples. 2 around 26 bar, one around 27 bar, one which we are not certain of its pressure but it is around 28 bar or a little higher, and one which had around 17\% liquid in the cell. All solid samples deviated from the empty cell background below 0.2K. An example of this behaviour for the 28 bar sample can be seen in Fig. \ref{fig:TripleP}. The sample containing 17\% liquid did not show this deviation. However, it did not follow the empty cell background above 0.2K, but rather had a curved profile with the period increasing with temperature.  
The data points of $\Delta p/p$ of the empty cell of all three modes have the same behaviour, and fall quite close to one another with a difference of up to 1x$10^{-6}$ at 0.7K. Here $\Delta p=p(T)-p(T=40mK)$. The data points are equilibrium points. At each step the temperature was held constant until the period stabilized, the error bars being the standard deviation of the spread at a constant period. For this particular experiment we were not able to measure stable data points at 30mK, but as the period hardly changes below 50mK, we take the 40mK data point as being close enough to a lower temperature value, and give a reasonable estimate of the signal size. 

\begin{figure}[]
\includegraphics[width=3.5in]{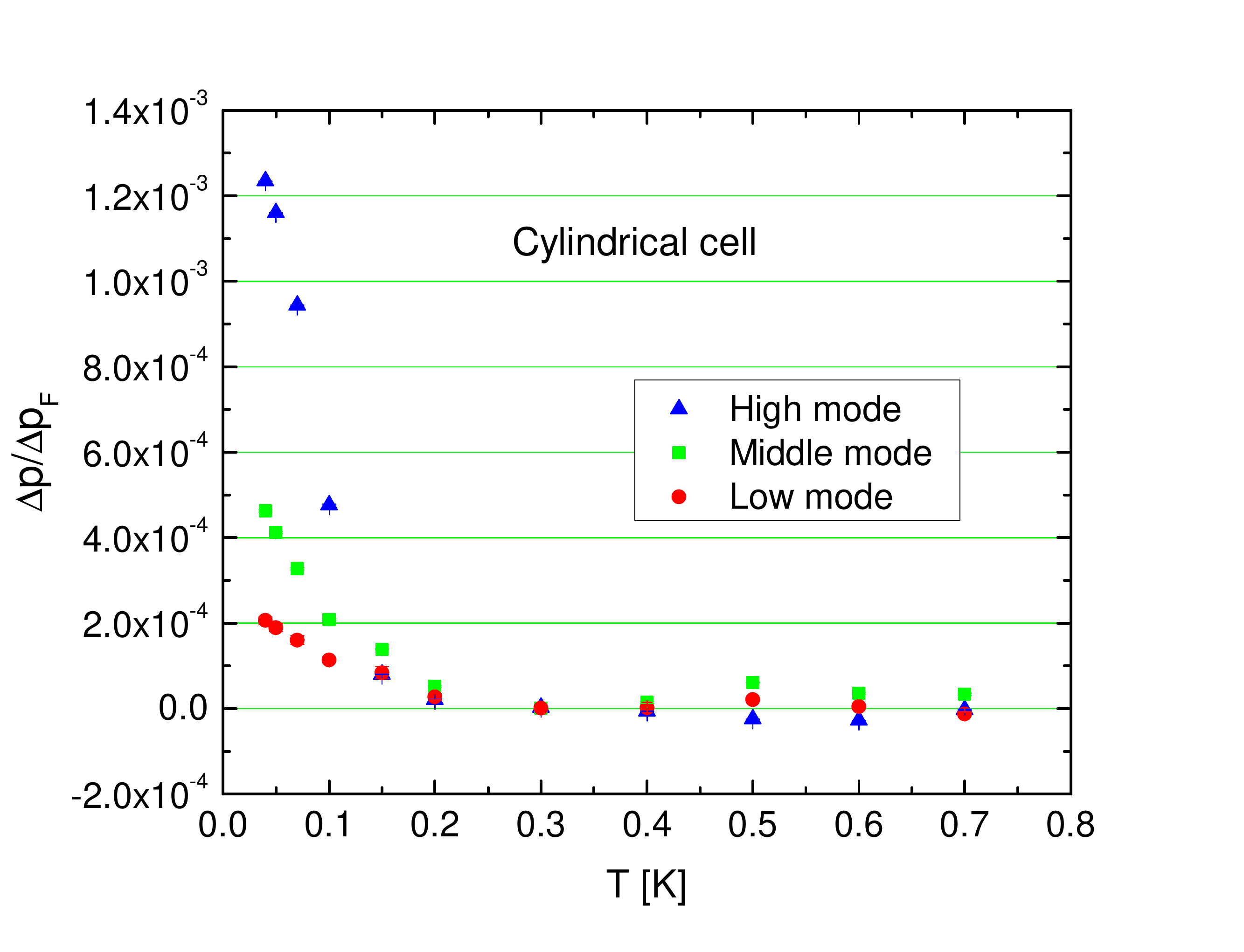} {}
\caption{(Color Online) Fractional period change for a sample with a pressure around 28bar.}
\label{fig:TripleDelP}
\end{figure}

The values of $\Delta p/\Delta p_F$ for the 28 bar sample are shown in Fig \ref{fig:TripleDelP}.  The error bars come from the small difference between the slopes of the empty and full cell data points. We matched the empty cell and full cell points at 0.3K. 

\begin{figure}[]
\includegraphics[width=3in]{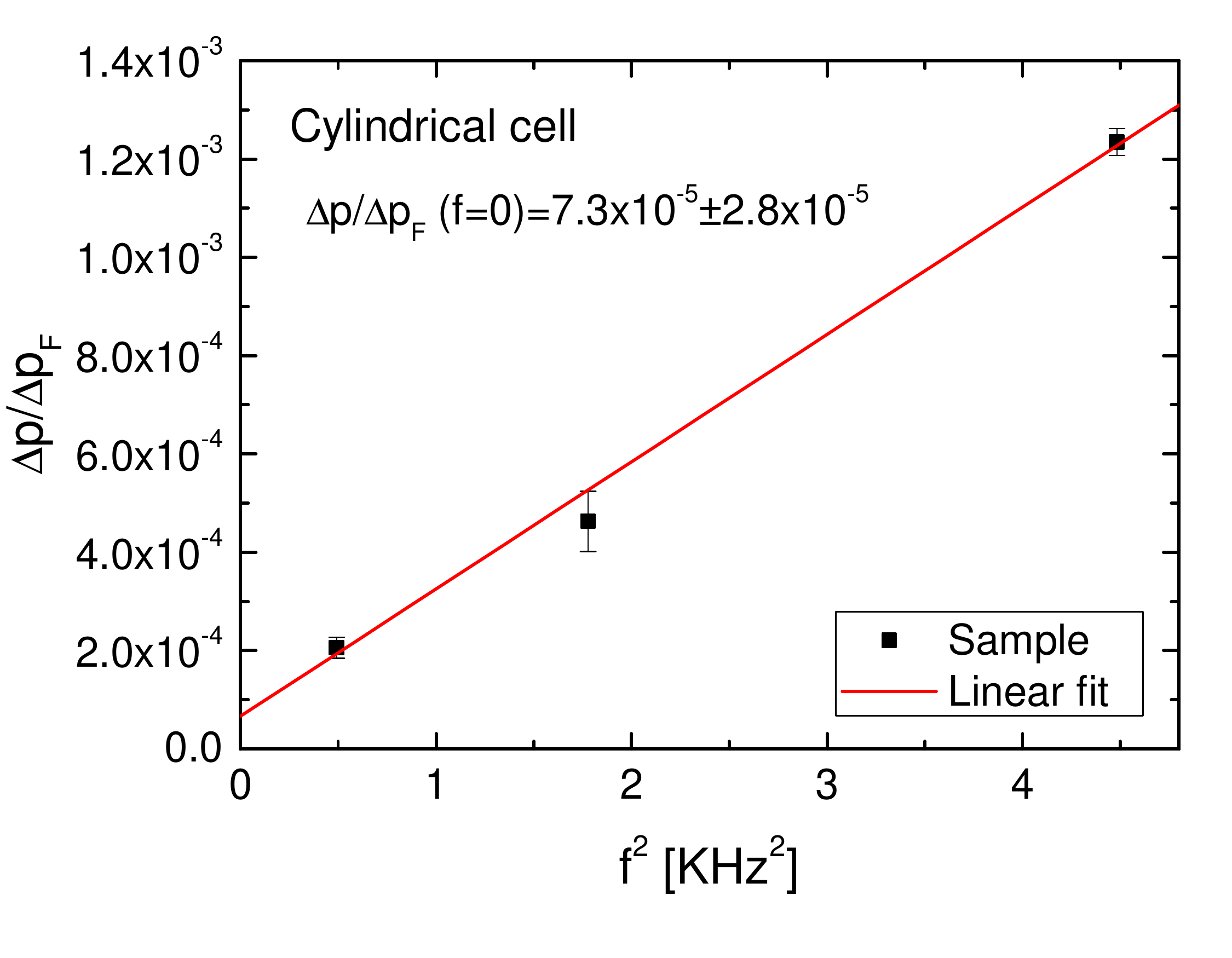} {}
\caption{(Color Online) FPS vs. f$^2$ for a 28 bar sample.}
\label{fig:TripleFPS}
\end{figure}

All our measurements were done with outer cell wall maximal velocities below 20 $\mu$m/sec, at a regime where $\Delta p/p$ changes by less than $10^{-7}$ by changing the velocity. The empty cell $\Delta p/p$ values for all three modes had almost the same dependence on the velocity of the cell, changing by about 4x$10^{-7}$ when increasing the velocities to 100 $\mu$m/sec.  The full cell period starts to deviate from the empty cell background at around 20 $\mu$m/sec.
The plot of the lowest temperature values (40mK) from our triple TO, shown in Fig. \ref{fig:TripleDelP}, vs. the frequency squared of each mode can be seen in Fig. \ref{fig:TripleFPS}. The line in the plot is a fit to a straight line. It gives a non-zero intercept of 7.3x$10^{-5}\pm$ 2.8x$10^{-5}$ with the y axis. The different intercepts for the rest of the samples varied between 3.55x$10^{-5}\pm$0.32x$10^{-5}$ and 1.33x$10^{-4}\pm$0.39x$10^{-4}$.

We preformed finite element method (FEM) simulations on our triple TO design as well, changing the shear modulus of the helium inside the cell. All shear modulus changes produced plots with zero y axis intercepts for the $\Delta p/\Delta p_F$ vs. $f^2$ plots. From these simulations, we find that a shear modulus change of around 20\% can account for the elastic part of our observed signal, but not for the frequency independent term. The only other frequency independent contribution which can exist in our cell is the filling line effect, which for our design for a 100\% change of the shear modulus of the helium in the fill lines gives $\Delta p/\Delta p_F$ of less than 3x$10^{-8}$ for all modes, three orders of magnitude smaller than our smallest non zero intercept.

		\subsubsection{Annular double mode oscillator}
We also performed measurements using an annular cell. A cross section of our double-frequency TO with an annular cell is shown in Fig. \ref{fig:Cells} (c). The cell, dummy piece and torsion rod were all constructed of Al 6061. The torsion rod had been heat treated. 
The different moments of inertia and torsion constants were:\\

\begin{tabular}{ l ||  l }

  Torsion constant  [dyn cm]  & Moment of inertia [g cm$^2$]\\
\hline
  $K_{cell}$= 1.69E+09  & $I_{cell}$=  26.0 \\
  $K_{dummy}$=1.35E+09  & $I_{dummy}$=  58.7  \\
 $K_{VI}$= 1.03E+9 & $I_{VI}$= 547.0 \\
  &  $I_{He}$= 0.249 \\
\end{tabular}
\\
\\

\begin{figure}[]
\includegraphics[width=3in]{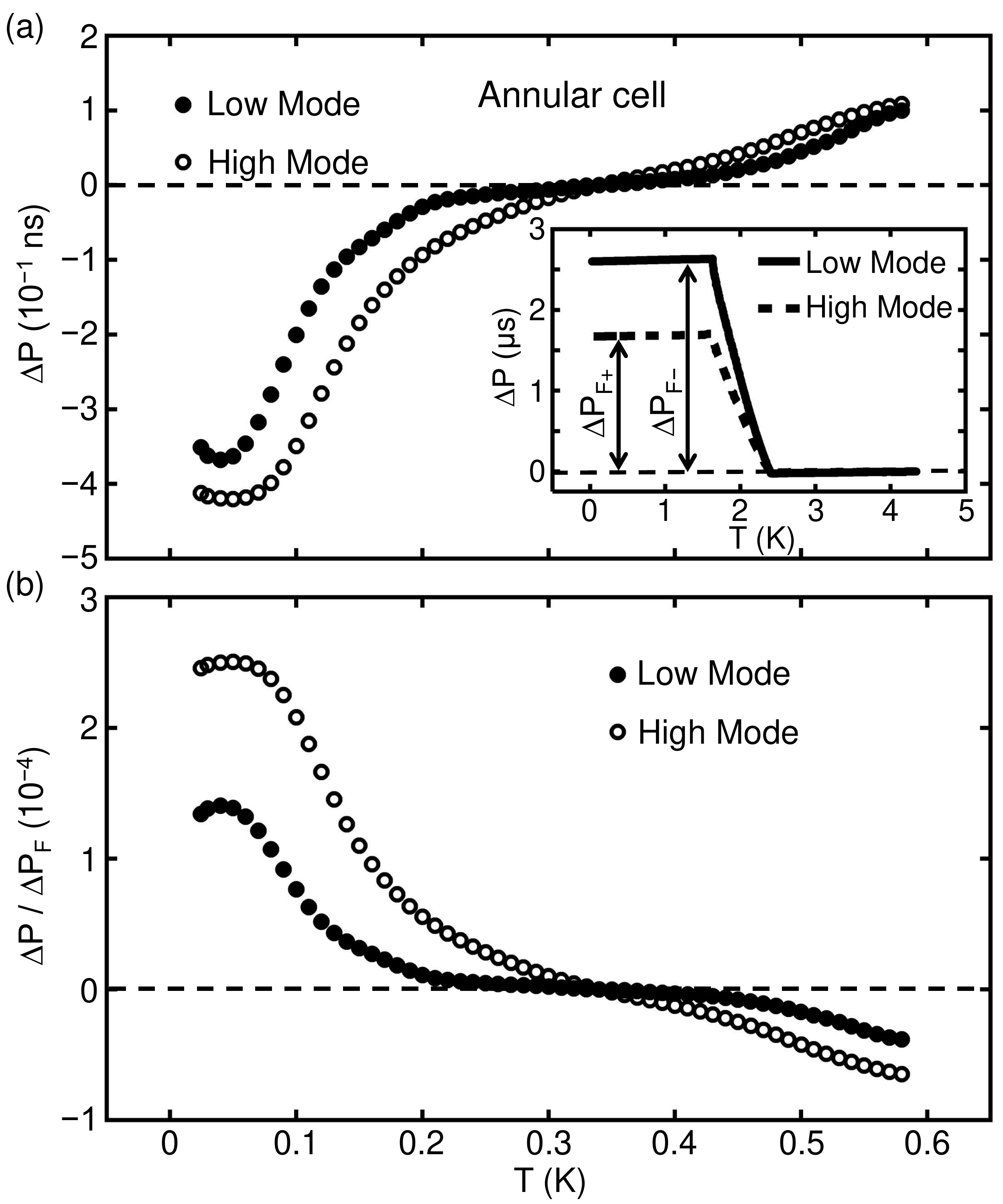} {}
\caption{(a) Period change vs. temperature for the two modes of the annular TO. Empty cell background subtracted. (b) The fractional period shift of the two modes. The inset of (a) shows the period change upon freezing a sample.}
\label{fig:AnnularPs}
\end{figure}

The torsion rods had inner fill-lines with a radius $r_{fill} = 0.017$ cm and an outer radius $r_{rod} = 0.256$ cm, for a ratio of $r_{fill} / r_{rod} = 0.067$.
At $T = 0.5$ K, the resonance periods of the empty TO were $p_- = 1.548$ ms for the low frequency $(-)$ mode and $p_+ = 0.648$ ms for the high frequency $(+)$ mode, with corresponding frequencies of $f_- = 646.0$ Hz and $f_+ = 1543.0$ Hz.
The measurements for the double TO were mainly temperature sweeps. We first cooled the TO to a base temperature of $0.02$ K, and after $p_\pm$ has fully relaxed to a constant value, we slowly warmed the TO at a fixed rate of $dT/dt$ = 0.33 mK/min and $p_\pm$ was recorded as function of temperature. 
The inset to Fig. \ref{fig:AnnularPs} (a) shows the period shift data for the two resonance modes as the sample is frozen. The total period shifts for the two modes, between the liquid and the solid phases, were $\Delta p_{F-} = 2.75$ $\mu$s and $\Delta p_{F+} = 1.69$ $\mu$s. These shifts determined the mass-loading sensitivities for the two modes. Based on a temperature of 1.63 K at which freezing ceases, the final sample pressure is estimated to be about 26 bar.
The temperature dependent values $p_\pm$ of the TO with this solid $^4$He sample inside were taken in two ways. First we measured $p_\pm$, starting from its fully relaxed equilibrium value at $T = 0.02$ K, at a constant warming rate of $dT/dt =$ 0.33 mK/min. We then cooled the sample back to $T = 0.02 $ and waited until $p_\pm$ relaxed back into its equilibrium value at 0.02 K. The sample was then warmed to a succession of fixed temperatures. At each fixed temperature, we waited until $p_\pm$ has fully relaxed to a constant. 
After the melting of the sample, we warmed the TO to $T = 20$ K and pumped the fill line overnight to evacuate any residual $^4$He which may remain inside the cell. The TO was then cooled down to $T = 0.02$ K and the empty cell values of $p_\pm$ were taken again in the two different ways described in the previous paragraph. The empty cell values of $p_\pm$ taken after the sample were very consistent with those taken before the sample. At temperatures between 0.02 K to 0.2 K, $p_\pm$ changed by a maximum of 0.01 ns. This yields an uncertainty of $5 \times 10^{-6}$ in the FPS induced by the solid $^4$He sample.
In Fig. \ref{fig:AnnularPs}(a), we extract data for the period shift $\Delta p (T)$, induced by solid $^4$He alone, at each mode as a function of temperature from $T$ = 0.02 K to $T$ = 0.58 K. Values of $\Delta p (T)$ are obtained by subtracting the average of two empty cell backgrounds from $p_\pm$ in the presence of solid $^4$He. 

\begin{figure}[]
\includegraphics[width=3in]{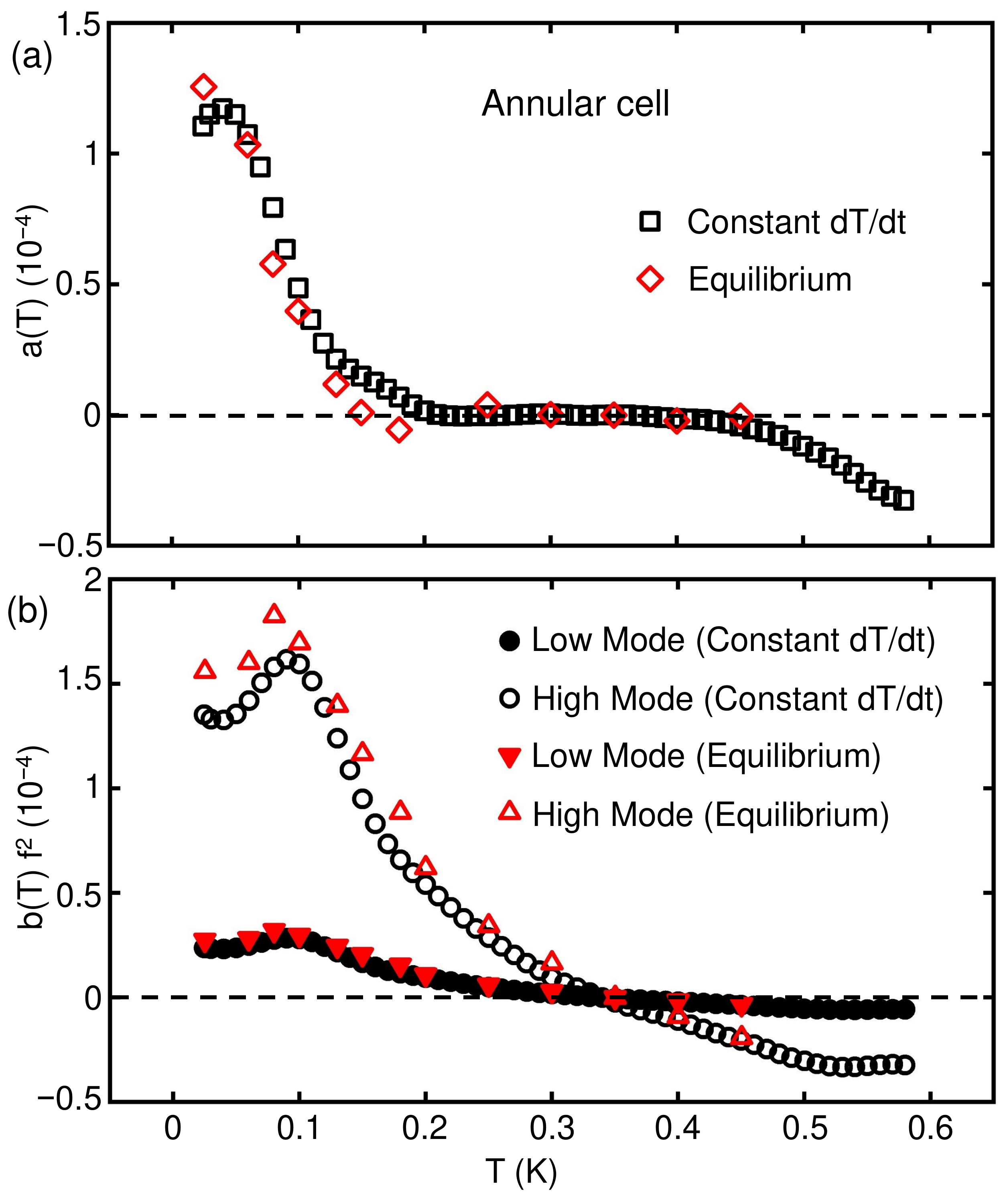} {}
\caption{(Color Online) The frequency independent (a) and elastic (b) parts of the signal. }
\label{fig:AnnularElastic}
\end{figure}

We observe clear anomalous period drops below 0.2 K where $\Delta p$ decreases rapidly. In the intermediate regime, $0.2 \text{ K} < T < 0.4 \text{ K}$, $\Delta p$ shows little variation over temperature. At $T > 0.4 \text{ K}$, a moderate increase in $\Delta p$ is observed which continues to higher temperatures. In Fig. \ref{fig:AnnularElastic} we show the elastic contributions to the different modes, b(T)f$^2$, and the frequency independent term, a(T), versus temperature. Both temperature sweeping and equilibrium data points are shown. The small discrepancy between the two can be a result of temperature lags in the sweeping data.  

As can be seen by Fig. \ref{fig:AnnularElastic}(a), in this annular cell, the same as in the open cylinder cells, we measured a frequency independent contribution to the data, on the order of 10$^{-4}$ of the total moment of inertia of the helium in the cell at the low temperature limit. The main difference in this cell being the difference in the elastic effect’s size. For the annular cell the elastic effect was about an order of magnitude smaller than in the open cylinders. Here too, as in Fig. \ref{fig:DoubleElastic} for the open cylinder cell, the frequency independent contribution seems to have a different temperature dependence than the elastic parts, leveling off around 150mK or so, whereas the elastic parts continue changing. 

		\subsection{The validity of our assumptions}

There are several assumptions underlying the analysis of our data.
The main one is that the elastic contribution to the signal
can be well approximated by an f$^2$ dependence. That term is actually
just the leading term in the series expansion of the changes in the effective moment of
inertia of the cell due to elastic effects \cite{Reppy_JLTP_Review2012}.

For our annular cell, using an approximation of an infinite cylinder with a radius $r_0 \gg \Delta r$, where $\Delta r$ is the width of the annulus   
\begin{multline}
\frac{I_{eff}}{I_{He}} \approx \frac{2}{\omega \Delta r} \sqrt{\frac{\mu}{\rho}} tan \biggl( \frac{1}{2} \omega \Delta r  \sqrt{\frac{\rho}{\mu}}\biggr) = \\
= 1+\frac{\Delta r ^2}{12} \frac{\rho}{\mu} \omega ^2 + \frac{\Delta r ^4}{120} \frac{\rho ^2}{\mu ^2} \omega^4 + ...
\end{multline}

Here $\rho$ and $\mu$ are the density and shear modulus of solid $^4$He.

We are interested in the change in this expression due to variations in the shear modulus $\Delta \mu = \mu _1 (T=30mK) - \mu _2 (T=300mK)$. 

\begin{multline}
\frac{\Delta I_{eff}}{I_{He}} = \\
=\frac{2}{\omega \Delta r} \biggl( \sqrt{\frac{\mu _1}{\rho}} tan \biggl( \frac{\omega \Delta r }{2}  \sqrt{\frac{\rho}{\mu _1}}\biggr)  - \sqrt{\frac{\mu _2}{\rho}} tan \biggl( \frac{\omega \Delta r }{2}  \sqrt{\frac{\rho}{\mu _2}}\biggr) \biggr) \\
= \frac{\Delta r ^2 \omega ^2 \rho }{12} \biggl( \frac{1}{\mu _1}-  \frac{1}{\mu _2} \biggr)+ \frac{\Delta r ^4 \rho ^2 \omega^4}{120} \biggl( \frac{1}{\mu _1^2}  -\frac{1}{\mu _2^2} \biggr) + ...
\end{multline}

The error introduced by considering only the first $\omega ^2$ term in our analysis is -
\begin{multline}
Error \biggl( \frac{\Delta I_{eff}}{I_{He}} \biggr) = \\
=1- \frac{\Delta r ^2 \omega ^2 \rho }{12} \biggl( \frac{\Delta \mu}{\mu _1 \mu _2} \biggr)  \bigg/ \frac{\Delta I_{eff}}{I_{He}}
\end{multline}

That gives, for the shear modulus values of $\mu _1 = 1.5 \times 10^8$ dyn cm$^{-2}$ and $\mu _2 = 1.0 \times 10^8$ dyn cm$^{-2}$ used in the appendix, a correction of 0.08\% for the high mode, and of 0.01\% to the low mode. 

For our open cylinder cell, with the infinite cylinder approximation 
\begin{multline}
\frac{I_{eff}}{I_{He}} \approx \frac{4 }{\omega r_0 } \sqrt{\frac{\mu}{\rho}} \: \frac{J_2 \biggl( r_0 \omega \sqrt{\frac{\rho}{\mu}} \biggr)}{J_1 \biggl( r_0 \omega \sqrt{\frac{\rho}{\mu}} \biggr)} = \\
= 1+\frac{r_0 ^2}{24} \frac{\rho}{\mu} \omega ^2 + \frac{r_0 ^4}{384} \frac{\rho ^2}{\mu ^2} \omega^4 + ...
\end{multline}
Where $J_n$(x) are the Bessel functions of the first kind.
Again, in our analysis, we consider only the $\omega ^2$ term in the series expansion for $ \frac{\Delta I_{eff}}{I_{He}}$.
Here, the errors we get are 1.9\% for the high mode and 0.3\% for the low mode. 
The shear modulus difference was evaluated from the size of the measured elastic contribution. We were using 
$\mu _1 = 1.5 \times 10^8$ dyn cm$^{-2}$ and $\mu _2 = 1.24 \times 10^8$ dyn cm$^{-2}$.
It is important to point out that the influence of the following terms, such as the f$^4$ correction, would be in 
expanding the horizontal axis of figures like Fig. \ref{fig:DoubleFPSvsf2}.
That, in turn, would slightly increase the zero frequency intercept.   

Another assumption of our model is that the shear modulus
of the helium is frequency independent. As can be seen in figures \ref{fig:BeamishShearModulus}  
and \ref{fig:BalibarShearModulus}, there is actually a strong frequency dependence of the shear modulus in the 
intermediate range of 50mK to 200mK, where the shear modulus itself changes with
temperature, however, for the frequencies used in our experiments, 
in the temperature ranges of either below about 40mK, and above around 250mK
the shear modulus has only a small dependence on frequency. From the polycrystalline
data by Beamish (shown in Fig. \ref{fig:BeamishShearModulus}), at 300mK, 
the difference in shear modulus for the frequencies
used in our experiment is less than 1\% of the total shear modulus. Hence, the correction to the frequency
independent term in the plots of FPS vs. f$^2$, where the signals are taken as the
difference in period between 300mK and 30mK (or 40mK for the triple TO), would be
less than 1\% of its value. It is worth mentioning that even in the intermediate 
range where the shear modulus changes the most with frequency, the correction to the data 
shown in Fig. \ref{fig:DoubleElastic} would be less than 2.7\% at most (around 0.1K)
in that range. 

Another issue is the strong dependence of the zero frequency intercept
on the measured value of the low mode. Therefore, the accuracy of the measurement
of that mode is crucial. Large mass loading sensitivities, as the ones
in our experiments, alongside low vibrational noise at the low mode's frequency enable
high sensitivity measurements with small errors.  

Lastly, we assumed a homogenous sample where the shear modulus of the helium in the cell would change by the
same amount all over the cell. That doesn't necessarily need to be the case,
but as it won't change the general frequency dependence, it will not have an effect
on the zero frequency term for our cells, where non f$^2$ contributions, such as the
fill line effect, were reduced to a minimum. 

	\section{Comparison with previous results} 
		\subsection{Previous multiple TOs}
Two other groups previously performed and reported the results of measurements of solid helium confined in a double TO, the Royal Holloway (RH) group \cite{Cowan} and the Rutgers group \cite{Kojima,KojimaPrb}. In these experiments, the distinction between the elastic and a possible supersolid effect is unclear, since the FPS signals contain significant contributions due to the stiffening of solid $^4$He in the torsion rods of both experiments.

In the Rutgers TO, the fill line drilled down the axis of the torsion rod to fill the cell had a diameter of 0.8mm, while the torsion rod itself had a diameter of 1.9mm. A 20\% change in the shear modulus of the helium in the torsion rod would give $\Delta p/\Delta p_F$ values of 1.1x10$^{-3}$ for the low mode and 0.86x10$^{-3}$ for the high mode. These numbers are comparable to the values reported in their 2007 paper \cite{Kojima}. In that case the fill line effect alone could account for the entire observed signal in the experiment, so any supersolid signal, should exist, would be very hard to subtract from the data. 

For the design of their cell the RH group chose a "pancake" geometry. This design is open to several elastic effects that contribute to the period shift signal. The cell had a relatively thin base plate, only 1 mm thick, for a radius of 7 mm. The holes drilled for filling their cells gave ratios of 0.17 and 0.37 for the inner to outer radii of the torsion rods. These enable elastic signals, such as the filling line effect and the Maris effect, on the order of their measured signals, making the distinction between a frequency independent contribution arising from elastic effects to one arising from a different origin difficult. 

In a previous work\cite{Mi_QFS2013}, Mi and Reppy reported the preliminary result from a double-frequency TO that was designed to be chiefly sensitive to elastic effect due to acceleration of solid $^4$He. In that experiment, a frequency-independent contribution to the FPS was observed that was equivalent to an inertial-mass-decoupling of the cylindrical sample corresponding to a NCRI/supersolid fraction of $1.2 \times 10^{-4}$. That number is comparable to the signals reported in this paper. 

		\subsection{Other relevant experiments}
There are only two other experiments showing signals smaller than the ones reported here. These are the rigid TO experiment by the Chan group \cite{ChanRigid} and a double oscillator experiment by the Kim group \cite{KimDouble}. In both these torsional oscillator experiments, in which they report an upper limit of 4x10$^{-6}$ to a possible supersolid signal, the experimental cell had a non-simply connected geometry, and the samples had pressures all above 34bar. It is possible that either the sample pressure has an effect on the size of the signal, or, should there be a percolating network allowing a flow to take place then the path length and cross section of the sample might play a role. 

Recent flow measurements done by 3 groups \cite{Hallock,BeamishFlow,ChanFlow} report flow in the helium which has a different temperature dependence than the observed signals in the torsional oscillators. The size of the flow is also orders of magnitude smaller than the effects detected in current TO experiments. 

		\subsection{Comparison with theory}
The original theoretical calculations \cite{Leggett} regarding the superfluid fraction of solid helium at low temperatures give an estimation of around 10$^{-4}$. Our numbers, ranging between 0.4x10$^{-4}$ and 1.3x10$^{-4}$ of the total moment of inertia of the solid helium in the cell, are comparable to the ones assumed by theoretical modeling of a supersolid \cite{Boronat2009}.
Path Integral Monte Carlo simulations results perhaps show no vacancy based supersolid in perfect single crystals containing several thousand atoms \cite{Svistunov2006}, yet don’t rule out other mechanisms for superflow in solid $^4$He, such as flow along percolated networks of dislocations or grain boundaries. 

\section{Conclusion}
Our results indicate the existence, for some, but not all, solid samples contained in a variety of sample geometries, of a frequency independent contribution to the period. The existence of such a signal can be taken, in the absence of another explanation, as evidence for the supersolid state.

	\section{Future tests}
There are yet some tests left to perform in order to pin down the exact nature of the apparent phenomenon – which will enable to state for certain if it is a supersolid or not. 
One of them is the blocked annulus. A supersolid phase would only give a period change in a torsional oscillator if the condensed phase has a closed circular path, so blocking its path should cause almost the entire signal to disappear.  Elastic effects, however, should not decrease by the same amount.  

Recent TO experiments \cite{ChanRigid} with solid $^3$He in the HCP phase have shown period shift signals similar to those seen in solid $^4$He. Again, the shear modulus of the HCP phase of solid $^3$He has an anomaly similar to that of solid $^4$He. which suggests that the $^3$He period shift signal must arise as a consequence of the elastic anomaly. Solid $^3$He is not Bose system and therefore not expected to exhibit a Bose-condensed supersolid phase. Therefore, if a frequency independent contribution were to be observed in the case of solid $^3$He, this would provide a strong argument against a supersolid origin for the observed zero frequency contribution seen in our $^4$He experiments. 

\begin{acknowledgements}
We acknowledge useful and encouraging discussions with P. W. Anderson, J. Beamish, W. F. Brinkman, M. H. W. Chan, J. C. Davis, D. A. Huse, and E. Mueller. This work was supported by the National Science Foundation through Grant DMR-060586 and CCMR Grant DMR-1120296, and partially funded by the New England Foundation, Technion. \\
\end{acknowledgements}

Author Information: anna.eyal@gmail.com

\bibliographystyle{spphys}       
\bibliography{supersolid}

\clearpage

\appendix
\section{Analytical Calculation}

In this appendix, we provide a more detailed discussion of the effects of shear modulus changes in solid $^4$He on the resonance periods of the annular torsional oscillator (TO) employed in this work. We approach the problem with an analytical calculation and a numerical simulation using Finite Element Method (FEM). The same type of analysis can be applied to any of our cells, as they were all designed to suppress elastic effects. In the following calculations it is assumed that the absolute value of the shear modulus of solid $^4$He is independent of frequency in the frequency range of our TO. Justification for this and other assumptions is discussed in section II D of the main text. 

\begin{figure}[]
\centering
\vspace{0.1in}
\includegraphics[width=2in]{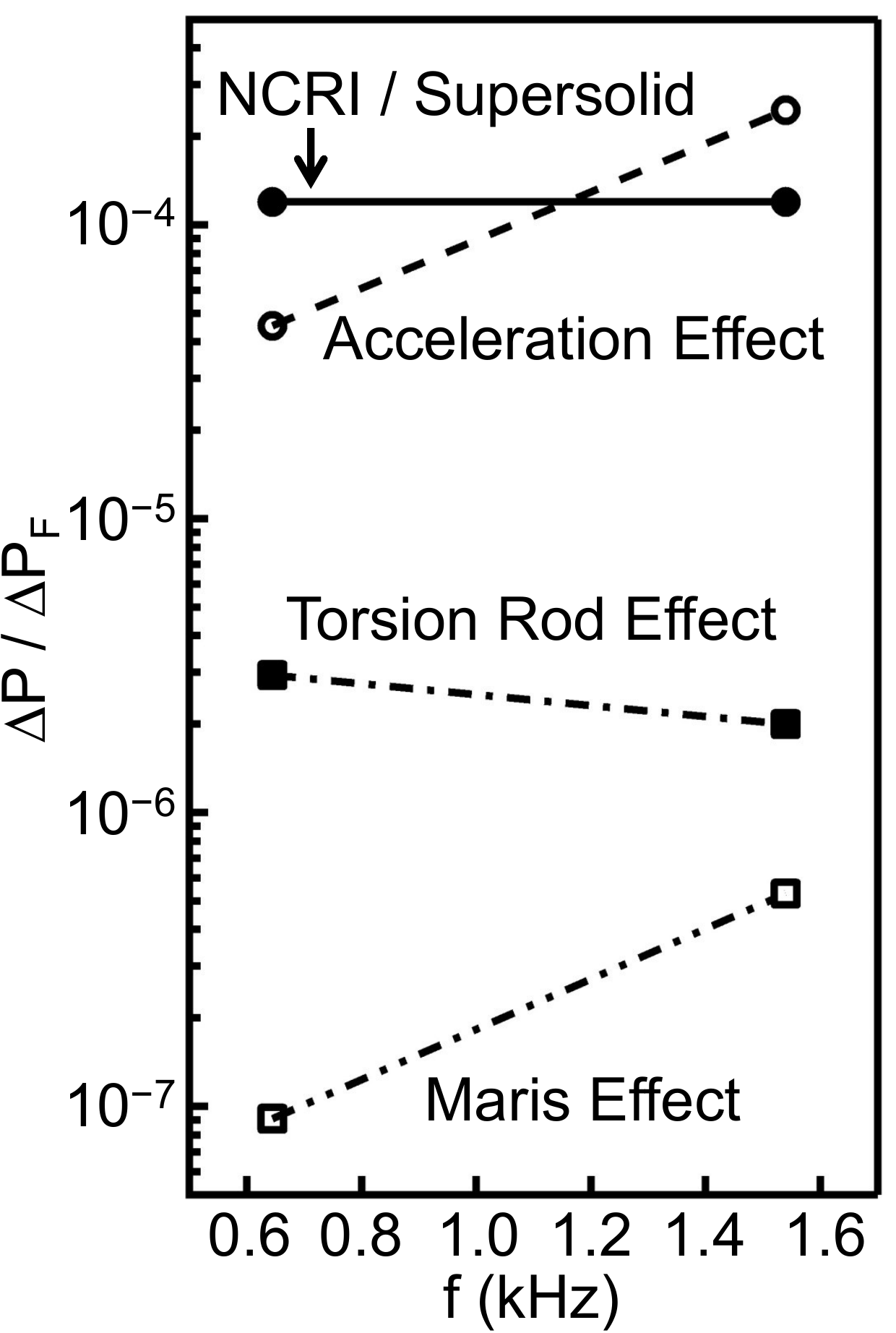} {}
\caption{Fractional period shifts for the annular cell shown in Fig. \ref{fig:Cells} (c) vs. their frequency for the different elastic contributions. Our measured frequency independent contribution is shown for comparison.}
\label{fig:FEMFPS}
\end{figure}

To visualize the frequency dependence for this particular cell, we plot the different contributions to the signal based on changes in $\mu$ in Fig.~\ref{fig:FEMFPS}. 

\subsection{Acceleration effect}

We first calculate the effect arising from the acceleration field of solid $^4$He. The majority of the solid $^4$He sample is confined in a long annular channel with inner radius $r_i = 0.635$ cm, outer radius $r_o = 0.794$ cm and height $L = 3.28$ cm. The remaining solid $^4$He sample is confined at the bottom of the sample volume in a thin cylindrical space with radius $r_c = r_i = 0.635$ cm and height $H = 0.127$ cm.

For the annular part of the sample, the amplitude of the displacement field, $\vec{u}$, as a function of radius from the symmetry axis of the sample $r$ is calculated to be \cite{Reppy_JLTP_Review2012}
\begin{equation}
\vec{u}(r) = \frac{r_m \theta_{0} \cos \left( (r - r_m) \omega \sqrt{\rho / \mu} \right)}{ \cos \left( \tfrac{1}{2} \Delta r \omega \sqrt{\rho / \mu} \right) } \vec{e}_{\theta}
\end{equation}
where $\Delta r = r_\text{o} - r_\text{i} = 0.159$ cm is the width of the annulus, $r_\text{m} = \tfrac{1}{2} (r_\text{o} + r_\text{i}) = 0.715$ cm is the mean radius of the annulus, $\omega = 2 \pi f$ where $f$ is TO frequency which takes on values of $f_-$ and $f_+$ at the two resonance modes, $\rho$ and $\mu$ are the density and shear modulus of solid $^4$He, $\theta_0$ is the maximum angular displacement in radians of the oscillating TO and $\vec{e}_{\theta}$ is the azimuthal direction defined with $z$-axis being the symmetry axis of the sample. In this estimate, we are neglecting the finite length of the annulus since it is much greater than the radius of the annulus, $L \gg r_\text{m}$. The first internal radial sound mode of solid $^4$He in this geometry occurs at a frequency of $f_s = \tfrac{1}{2 \Delta r} \sqrt{\mu / \rho} = 86.3$ kHz, nearly two orders of magnitude higher than the TO frequencies. We are therefore justified in expanding $\vec{u}(r)$ in terms of $\omega$ and keeping only the two lowest order terms.

To convert the displacement field into a back-action torque $\tau_\text{He}$ on the TO, we integrate the $\theta$ component of the displacement field according to $\tau_\text{He} = \int_{r_\text{i}}^{r_\text{o}} dr \rho \omega^2 u_\theta (r)$ \cite{Reppy_JLTP_Review2012,Graf_Glassy_DoubleTO}. The effective moment of inertia of solid $^4$He is then
\begin{equation}
I_\text{eff} = \tau_\text{He} / (\theta_0 \omega^2) \approx I_\text{He} \left( 1 + \frac{1}{2} (\Delta r)^2 \omega^2 \frac{\rho}{\mu} \right)
\end{equation}
where $I_\text{He} = \tfrac{1}{2} \rho L \pi (r_\text{o}^4 - r_\text{i}^4) = 0.242$ g$\,$cm$^{2}$ is the moment of inertia of the annular part of the $^4$He sample. An increase in $^4$He shear modulus decreases $I_\text{eff}$ by an amount $\Delta I_\text{eff} \propto \omega^2$, which leads to a fractional period shift (FPS), $\Delta P / \Delta P_\text{F} = \Delta I_\text{eff} / I_\text{eff}$. Therefore, the observed FPS is proportional to $f^2$. The cylindrical part of the sample is treated in an analogous manner. The effective moment of inertia has a similar form \cite{Reppy_JLTP_Review2012}:
\begin{equation}
I_\text{eff} \approx I_\text{He} \left( 1 + \frac{1}{2} H^2 \omega^2 \frac{\rho}{\mu} \right)
\end{equation}
The moment of inertia of this part of the sample is $I_\text{He} = \frac{1}{2} \rho H \pi r_\text{c}^4 = 0.0065$ g$\,$cm$^2$.

To estimate the magnitude of FPS at each resonance mode due to shear-stiffening of solid $^4$He, we vary $\mu$ from $1.5 \times 10^8$ dyn$\,$cm$^{-2}$ to $1 \times 10^8$ dyn$\,$cm$^{-2}$ and compute the fractional change in the total effective moment of inertia which is the sum of expressions in equations (A2) and (A3). The result is a FPS of $1.42 \times 10^{-4}$ for the high frequency mode and $0.25 \times 10^{-4}$ for the low frequency mode, both being very close to the elastic contributions to our observed FPS. We note that we have ignored the finite shear modulus of the aluminum alloy constituting the TO. The next effect we discuss takes the elasticity of the cylindrical wall of the cell into account. In the FEM computation presented in the next section, the finite shear modulus of the entire TO is taken into account which amplifies the FPS calculated here, although the FPS remains proportional to f$^2$.

\subsection{Twisting of the cell walls}
In Ref. \cite{Reppy_JLTP_Review2012}, we discussed a correction to the TO periods due to the twisting of the cell walls. This effect arises from the fact that the bottom of the cell undergoes a slightly larger angular displacement than the top of the cell, due to the elastic displacement of the cylindrical cell wall. The size of this effect is approximated by changes in the effective moment of inertia $I_\text{E}$ of the cell itself, which is given by
\begin{equation}
I_\text{E} \approx I_\text{U} \left( 1 + \frac{1}{3} U^2 \omega^2  \frac{\bar{\rho}}{\bar{\mu}} \right)
\end{equation}
where $I_\text{U} = 7.05$ g$\,$cm$^2$ is the moment of inertia of the cell from the bottom of the solid $^4$He sample to the top, $U = 3.43$ cm is the total length of this part of the cell, $\bar{\rho}$ and $\bar{\mu}$ are the weighted averages of the density and shear modulus of the cell$ + ^4$He system. Specifically for our cell, since $I_\text{U} \gg I_\text{He}$, $\bar{\rho} \approx \rho_\text{al}$ where $\rho_\text{al}$ = 2.7 g$\,$cm$^{-3}$ is the density of aluminum. $\bar{\mu}$ is given by considering the cross-section of the sample space. At radius $r < 0.635$ cm and $0.794$ cm $ < r <$ 0.921 cm, the cross-section is occupied with aluminum. Solid $^4$He only occupies the annular region 0.635 cm $ < r <$ 0.794 cm. Approximating the cell as a torsion rod with this given cross-section, we calculate that the fractional contribution of solid $^4$He to the total torsion constant is $0.48 \mu / \mu_\text{al}$ where we used $\mu_\text{al} = 2.7 \times 10^{11}$ dyn$\,$cm$^{-2}$ for the shear modulus of the Al 6061 alloy. This suggests that $\bar{\mu} \approx \mu_\text{al} + 0.48 \mu$. 

As before, we vary $\mu$ from $1.5 \times 10^8$ dyn$\,$cm$^{-2}$ to $1 \times 10^8$ dyn$\,$cm$^{-2}$, obtaining a change in $I_\text{E}$ of $\Delta I_{\text{E}-} = 4.05 \times 10^{-7}$ g$\,$cm$^{2}$ for the low frequency mode and $\Delta I_{\text{E}+} = 2.31 \times 10^{-6}$ g$\,$cm$^{2}$ for the high frequency mode. Normalizing these values by the moment of inertia of solid $^4$He, $I_\text{He}$, the FPS is $1.63 \times 10^{-6}$ for the low frequency mode and $9.27 \times 10^{-6}$ for the high frequency mode. We note that this effect has the same $f^2$ dependence as the acceleration effect. Comparing the values of the FPS calculated here to those calculated for the acceleration effect, we see the finite shear modulus of the cylindrical cell wall alone enhances the acceleration effect by about 7$\%$.

\subsection{Torsion rod or fill line effect}
The effect on TO periods produced by the stiffening of solid $^4$He inside the fill-line drilled through the torsion rods has been discussed by Beamish el al. \cite{Beamish_TorsionRodEffect}. The expression for the period shift in the case of a single mode TO where the diameter of the fill line is much smaller than the diameter of the torsion rod is
\begin{equation}
\frac{\Delta p}{ p}= - \frac{\Delta \mu_{He}} { 2 \mu_{rod} }\frac{ r_{fill}^4}{r_{rod}^4}
\end{equation}
Thus, the effect can be minimized by reduction in the radius of the fill line $r_{fill}$.
When the $^4$He freezes in the fill line the shear modulus of the $^4$He increases from zero for the liquid, to about 1.5 x 10$^8$ dyn/cm$^2$ for the solid and the reduction in period due to this freezing is given by the formula above, with $\Delta \mu_{He}=1.5 \text{x} 10^8$.  
The freezing of the sample also produces a change in the TO, 
\begin{equation}
\frac{\Delta p_F}{p}=-\frac{I_{He}}{2 I_{cell}}
\end{equation}
Then in the case of the single mode TO, the contribution to the fractional period shift due to the freezing of the $^4$He in the fill line is given by
\begin{equation}
FPS=\frac{\Delta p }{\Delta p_F} =\frac{ \Delta \mu_{He} }{ \mu_{rod}} \frac{I_{cell}}{I_{He}} \frac{r_{fill}^4}{r_{rod}^4}
\end{equation}
In the case of double or triple mode TOs, multiple torsion rods are involved and the period shifts are obtained from the equations of motion.
For our annular cell's torsion rods, which have a shear modulus of $\mu_{rod}=\mu_\text{Al}$, the FPS produced by a change in solid $^4$He shear modulus $\Delta \mu$ is \cite{Beamish_TorsionRodEffect}, in the limit of $r_\text{fill} \ll r_\text{rod}$, 
\begin{equation}
\frac{\Delta p_\pm}{\Delta p_{F\pm}} \approx \frac{p_\pm}{2 \Delta p_{F\pm}} \frac{\Delta \mu_{He}}{\mu_\text{Al}} \left( \frac{r_\text{fill}}{r_\text{rod}} \right)^4
\end{equation}
For that TO, $r_\text{fill} / r_\text{rod} = 0.067$, and the measured mass-loading sensitivities are $\Delta p_{F-} = 2.75$ $\mu$s and $\Delta p_{F+} = 1.69$ $\mu$s. 
Taking the solidification of the helium in the fill line as an upper limit on this effect, we can use 
a shear modulus change of 100\% for the solid $^4$He sample, so that $\Delta \mu = 1.5 \times 10^8$ dyn$\,$cm$^{-2}$. That gives a FPS of $3.15 \times 10^{-6}$ for the low frequency mode and $2.15 \times 10^{-6}$ for the high frequency mode. Since these estimates are two orders of magnitude smaller than the measured signals, we conclude that changes in the shear modulus of the solid $^4$He inside the torsion rods form a negligible contribution to the observed FPS.

\subsection{Maris effect}
A subtle effect was discussed by H.~J.~Maris \cite{Maris_Effect} where the solid $^4$He sample inside the TO cell modifies the effective torsion constant of the torsion rod by exerting a counter-stress on the elastic base plate of the sample volume connected to the torsion rod. The magnitude of this ``Maris Effect'' depends on the stiffness of the base plate relative to the solid $^4$He sample in contact with the plate, and can be diminished by an increase in thickness or stiffness of the base plate. The effect is more significant for cylindrical sample geometries than for annular cell geometries. To estimate an upper bound on the size of the Maris effect for our annular cell, we replace the inner Al cylinder of the cell by solid $^4$He, so that the solid $^4$He sample is a cylinder with height $L = 3.28$ cm and radius $r_\text{o} = 0.794$ cm. The result from this simplification is an upper bound on the magnitudes of the signals if the actual geometry is used.

Following the approach outlined in Ref.~\cite{Maris_Effect}, we calculate that for a change of $\Delta \mu$ in solid $^4$He shear modulus, the corresponding change in the torsion constant $k_\text{c}$ of the cell $\Delta k_\text{c}$ is
\begin{equation}
\Delta k_\text{c} = 3.2 \times 10^{-9} (\Delta \mu / \mu) k_\text{c}
\end{equation}
It is noteworthy that the Maris effect only affects the effective torsion constant of the rod attached to the cell, whereas the torsion rod effect affects the torsion constants of both the cell and the dummy oscillator torsion rods. Consequently, the frequency dependences of the two effects are different.

The induced period shifts $\Delta p_\pm$ are obtained by increasing $k_\text{c}$ by an amount $\Delta k_\text{c}$ and calculating the decrease in the TO periods,
\begin{multline}
  p_\pm = 2 \pi \biggl[\frac{I_\text{c} \left( k_\text{c} +  k_\text{d} \right) +I_\text{d} k_\text{c}}{2 I_\text{c} I_\text{d}} \\
\biggl( 1 \pm \sqrt{1 - \frac{4 I_\text{c} I_\text{d} k_\text{c} k_\text{d}}{(I_\text{c} \left( k_\text{c} +  k_\text{d} \right) +I_\text{d} k_\text{c})^2}} \biggr) \biggr]^{- 1/2}
\end{multline}
For a 100$\%$ change in $\mu$, the period shifts are $\Delta p_- = 2.5 \times 10^{-4}$ ns and $\Delta p_+ = 9.0 \times 10^{-4}$ ns. These values give FPS of $9.1 \times 10^{-8}$ for the low frequency mode and $5.3 \times 10^{-7}$ for the high frequency mode, both being more than two orders of magnitude smaller than the observed FPS.

\subsection{Dissipation of solid $^4$He}
The additional dissipation introduced by the solid $^4$He in our experiment is very small. The increase in $1/Q$ where $Q$ is the mechanical qualify factor of the TO after the cell is filled with solid $^4$He has a peak value of only $5 \times 10^{-8}$ around a temperature of 0.1 K for both modes. The shifts in resonance periods associated with such changes in dissipation are $\Delta p_- = 3.87 \times 10^{-9}$ ns and $\Delta p_+ = 1.62 \times 10^{-9}$ ns. These values correspond to FPS of $1.41 \times 10^{-12}$ for the low frequency mode and $0.96 \times 10^{-12}$, eight orders of magnitude smaller than the FPS measured in the experiments.

\subsection{FEM computations}

\begin{figure}
	\centering
	\includegraphics[width=\columnwidth]{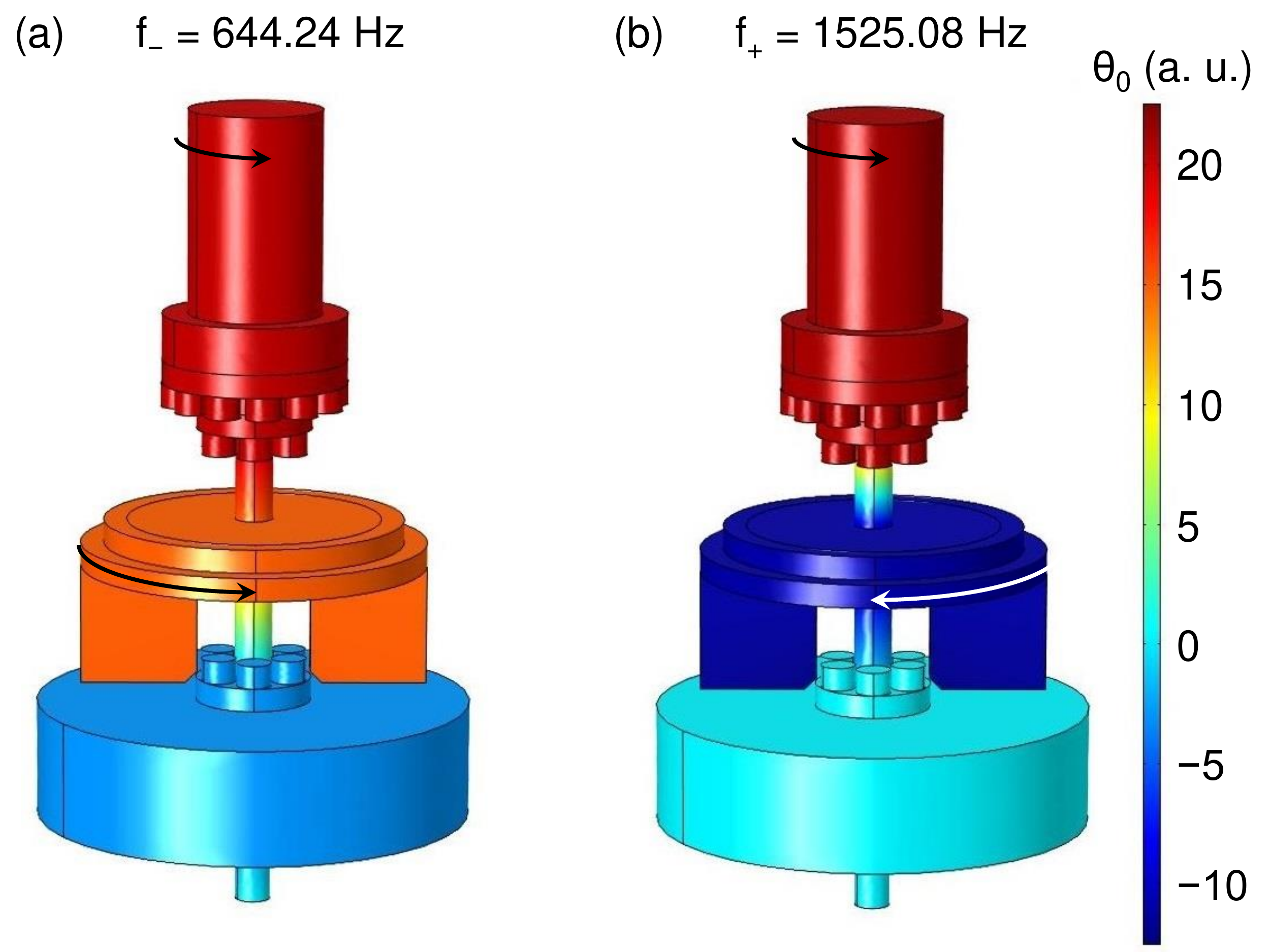}
	\caption{(Color Online) FEM simulations of the amplitude of angular displacements $\theta_0$ and resonance frequencies $f_\pm$ of the two resonance modes of the annular TO used in this work. All dimensions used in building the FEM model are the measured values of the actual apparatus. $\theta_0$ is plotted in arbitrary units. The bottom oscillator is the vibration isolator.}
	\label{fig:FEM}
\end{figure}

FEM calculations of TO resonance periods are performed with commercial software package COMSOL Multiphysics (structural mechanics module, COMSOL Multiphysics v4.3b, COMSOL Inc., 2013). A mesh consisting of 20745 domain elements is created, and the program solves for the eigen-frequencies $\omega$ of the Navier-Cauchy equation for the amplitude of the displacement field, $\vec{u}$, of the entire TO:
\begin{equation}
- \rho \omega^2 \vec{u} - \nabla \cdot \bar{\bar\sigma} = 0
\end{equation}
where $\bar{\bar\sigma}$ is the stress tensor. A plot for the amplitude of angular displacement, $\theta_0$, at the two resonance modes is shown in Fig. \ref{fig:FEM}. We have included a vibration isolator in the model for added precision. It can be seen that the values of $\theta_0$ have the same sign at the cell and the dummy oscillator for the low frequency mode, but opposite signs at the high frequency mode. Hence the two modes have shapes expected from analytical calculations, where the cell and the dummy oscillator rotate in-phase at the low frequency mode and out-of-phase at the high frequency mode. From the values of $\theta_0$, we also extract the scale factors $D_\pm$ relating the angular velocity of the dummy oscillator $\dot{\theta}_\text{d}$ to that of the cell $\dot{\theta}_\text{c}$, $\dot{\theta}_\text{c} = D_\pm \dot{\theta}_\text{d}$, which turns out to be $D_- = 1.40$ and $D_+ = -1.65$. From Fig.~\ref{fig:FEM}, one can see the eigen-frequencies calculated by the program match those measured experimentally very closely. We also check for the accuracy of the model by varying the density of solid helium $\rho$ from 0 to 0.2 g$\,$cm$^{-3}$. The calculated shifts in resonance periods are 2.76 $\mu$s and 1.72 $\mu$s for the low and high frequency modes respectively, again matching the experimental values $\Delta p_{F-}$ and $\Delta p_{F+}$.

To study the effect of changing the solid $^4$He shear modulus $\mu$, we shift the value of $\mu$ from $1.5 \times 10^8$ dyn$\,$cm$^{-2}$ to $3 \times 10^8$ dyn$\,$cm$^{-2}$ and calculate the shifts in periods at the two modes, $\Delta p_\pm$. Normalizing these shifts by the mass-loading values $\Delta p_{F \pm}$, the calculated FPS are $0.454 \times 10^{-4}$ for the low frequency mode and $2.46 \times 10^{-4}$ for the high frequency mode. These values are likely to be quantitatively more accurate than the analytical approach, since the finite shear modulus of the entire TO is taken into account, and so are the specifics of the cell geometry.

An important message imparted by the FEM simulation is the frequency dependence of the signals produced by the changing shear modulus of solid $^4$He. The ratio of the computed FPS at the two modes is $2.46/0.454=(f_+/ f_-)^{1.96}$. The exponent of 1.96 is in excellent agreement with the analytical prediction of 2, suggesting that changing solid $^4$He shear modulus indeed produces a FPS that is proportional to $f^2$ for the annular cell presented in this work. We repeated the FEM analysis with all our cells, and got the same $f^2$ dependence for the elastic effects in all of them.

\end{document}